\documentclass[journal]{IEEEtran}

\usepackage{amssymb}
\usepackage{amsbsy}
\usepackage[cmex10]{amsmath}
\usepackage{stfloats}
\usepackage{graphicx}
\usepackage{epstopdf}
\usepackage{subfigure}
\usepackage{tabularx}
\usepackage{epsfig,epsf,color,balance,cite}
\usepackage{bm}
\usepackage{multirow}
\usepackage{array}
\usepackage{setspace}
\usepackage{enumerate}
\usepackage{booktabs}
\usepackage{footnote}

\begin{document}

\title{Interference Management by Harnessing Multi-Domain Resources in Spectrum-Sharing Aided Satellite-Ground Integrated Networks}

\author{Xiaojin Ding,~\IEEEmembership{Member,~IEEE}, Yue Lei, Yulong Zou,~\IEEEmembership{Senior Member,~IEEE}, Gengxin Zhang,~\IEEEmembership{Member,~IEEE} and Lajos Hanzo,~\IEEEmembership{Life Fellow,~IEEE}% <-this % stops a space

\thanks{\hspace{1em}This work presented was partially supported by the National Science Foundation of China (No. 62171234, U21A20450, 91738201), the Jiangsu Province Basic Research Project (No. BK20192002), and the China Postdoctoral Science Foundation (No. 2018M632347).}

\thanks{Xiaojin Ding, Yue Lei, Yulong Zou and Gengxin Zhang are with the Telecommunication and Networks National Engineering Research Center, Nanjing University of Posts and Telecommunications, Nanjing 210003, China. E-mail: \{dxj, 1022072005, yulong.zou, zgx\}@njupt.edu.cn.}

\thanks{L. Hanzo is with the Department of Electronics and Computer Science, University of Southampton, Southampton, United Kingdom. (E-mail: lh@ecs.soton.ac.uk)}

\thanks{Y. Zou would like to acknowledge the financial support of the National Natural Science Foundation of China under Grant 62271268, the Jiangsu Provincial Key Research and Development Program under Grant BE2022800, and the Jiangsu Provincial 333 Talent Project.}

\thanks{L. Hanzo would like to acknowledge the financial support of the Engineering and Physical Sciences Research Council projects EP/W016605/1, EP/X01228X/1 and EP/Y026721/1 as well as of the European Research Council's Advanced Fellow Grant QuantCom (Grant No. 789028)}
}

\maketitle
\begin{abstract}
A spectrum-sharing satellite-ground integrated network is conceived, consisting of a pair of non-geostationary orbit (NGSO) constellations and multiple terrestrial base stations, which impose the co-frequency interference (CFI) on each other. The CFI may increase upon increasing the number of satellites. To manage the potentially severe interference, we propose to rely on joint multi-domain resource aided interference management (JMDR-IM). Specifically, the coverage overlap of the constellations considered is analyzed. Then, multi-domain resources - including both the beam-domain and power-domain - are jointly utilized for managing the CFI in an overlapping coverage region. This joint resource utilization is performed by relying on our specifically designed beam-shut-off and switching based beam scheduling, as well as on long short-term memory based joint autoregressive moving average assisted deep Q network aided power scheduling. Moreover, the outage probability (OP) of the proposed JMDR-IM scheme is derived, and the asymptotic analysis of the OP is also provided. Our performance evaluations demonstrate the superiority of the proposed JMDR-IM scheme in terms of its increased throughput and reduced OP.
\end{abstract}


\IEEEpeerreviewmaketitle

\section{Introduction}
\IEEEPARstart{L}{ow-}{Earth-Orbit (LEO) satellites have the potential of extending the coverage of terrestrial networks, while maintaining a low propagation delay \cite{BIB01}. Moreover, deploying base stations onboard of LEO satellites is capable of promptly setting up a networking infrastructure even in natural disaster areas. Thus, LEO satellites have attracted extensive research attention all over the world. Numerous LEO constellations have been proposed, as exemplified by Telesat \cite{BIB02}, Starlink \cite{BIB03}, and OneWeb \cite{BIB04}. These constellations are capable of simultaneously providing multiple LEO satellite links for a single user \cite{BIB05,BIB06}. Therefore, the power constraint of conventional small constellations can be readily overcome, since the required power can be jointly provided by multiple satellites. However, multiple satellites simultaneously occupy the spectral resources of a constellation, hence reducing the spectral efficiency.

Recently, sophisticated spectrum sharing techniques have emerged as an effective means of addressing the spectrum scarcity problem \cite{BIB07}. The most popular spectrum sharing modes are the overlay \cite{BIB08} and the underlay mode \cite{BIB09}. {\em In the overlay mode}, spectrum sharing is performed with the aid of accurate spectrum sensing and spectrum prediction both in integrated satellite-terrestrial networks \cite{BIB08,BIB10}, as well as in integrated geostationary (GEO) and nongeostationary (NGSO) scenarios \cite{BIBour}. Since spectrum sensing may fail to work reliably, the potential blind spots of spectrum sensing were explored for assisting spectrum sharing between GEO and NGSO scenarios in \cite{BIB11}. These overlay based spectrum-sharing schemes beneficially exploit the existence of the spectrum holes. By contrast, spectrum sharing may also be exploited even in the absence of spectrum holes with the aid of the underlay mode, albeit at the cost of incurring co-frequency interference~\cite{BIB12}, which has to be managed. {\em Hence the underlay mode tends to require a combination of the following interference-management techniques:} accurate power control \cite{BIB13}, frequency hopping \cite{BIB14}, sophisticated channel selection \cite{BIB15}, some frequency-reuse planning \cite{BIB16}, and refined beam control \cite{BIB17}. Specifically, a pair of optimal power control schemes were designed for a LEO satellite constellation in support of coexistence with terrestrial networks in \cite{BIB13}. Frequency hopping was used for spectrum sharing between fixed satellite services and tactical data links in \cite{BIB14}. As a further advance, a game-theory aided channel selection method was designed to meet the access requirements of primary and secondary users in \cite{BIB15}. To perform spectrum sharing for both cellular, as well as NGSO and GEO systems, the required geographic protection/exclusion-zone of frequency reuse was calculated in \cite{BIB16}. Furthermore, adaptive beam control was proposed for managing the interference imposed on the terrestrial networks by satellites \cite{BIB17}.

In contrast to the aforementioned single-dimensional spectrum-sharing schemes used in the overlay mode, multi-dimensional scheduling aided spectrum sharing schemes have also been designed for alleviating the co-frequency interference (CFI) \cite{BIB18,BIB19,BIB20}. A joint resource allocation algorithm was designed for increasing the throughput \cite{BIB18} by considering the interbeam interference, as well as the bandwidth allocation and power allocation. Moreover, sophisticated user association, bandwidth assignment and power allocation were jointly used for spectrum sharing in a terrestrial-satellite network \cite{BIB19}. As a further development, a joint beamforming and power allocation scheme was designed for allowing a satellite network to share its spectrum with a terrestrial network \cite{BIB20}. For a coexisting LEO and GEO satellite system, a joint power allocation supported by tight cooperation amongst the LEO satellites was proposed for managing the interference imposed on GEO users by the LEO satellites.

Whilst extensive efforts have been invested in supporting the coexistence of GEO, LEO and terrestrial wireless networks (CGLT), less attention has been dedicated to satellite-ground integrated networks (SGIN) relying on multiple NGSO constellations and terrestrial systems. In contrast to the CGLT scenarios, spectrum-sharing aided NGSO constellations travel at speed, hence resulting in violently time-varying interference. Moreover, NGSO constellations in each other's vicinity may impose severe CFI on each other's receivers in the presence of spectrum sharing, which may be classified into intra-constellation and inter-constellation (system) interference. Compared to the inter-constellation CFI, it is less challenging to manage the intra-constellation CFI, since the owner has the ability to configure its constellation \cite{BIB22}. To avoid the inter-constellation CFI, the inter-system interference of NGSO constellations in coexistence with GEO systems was analyzed in \cite{BIB22}, and the efficiency of low-complexity interference management schemes was also evaluated. However, the performance evaluations in \cite{BIB22} indicated that the existing spectrum regulation schemes may be unable to guarantee the target-reliability of the system analyzed. Moreover, the interference scenario of NGSO satellite communication systems was analyzed, and the challenges of alleviating the CFI between NGSO constellations were investigated in \cite{BIBNGSOCFI}.

Against the above backdrop, we conceive inter-system interference management techniques for spectrum-sharing aided SGIN, including a pair NGSO constellations and multiple terrestrial base stations (BSs). Explicitly, one of the NGSO constellations, namely NGSO 1, shares its spectrum both with the other constellation, denoted by NGSO 2, and with the BSs. Our interference analysis indicates the existence of severe CFI. To manage the CFI caused by spectrum sharing, an optimization problem is formulated for maximizing the throughput of NGSO 1, whilst still meeting the transmission requirements of both NGSO 2 and of the BSs. However, it is challenging to directly solve this complex problem, since it depends on numerous factors, such as the coverage overlap of NGSO constellations, the transmit power, on the number of satellites available, the tele-traffic model, the channel state information (CSI) and on the transmission requirements. Thus, this excessively complex problem is decoupled. Specifically, we first analyze the coverage overlap between NGSO 1 and NGSO 2. To further manage the CFI, the beam-domain and power-domain resources are jointly scheduled for solving the optimization problem formulated. Additionally, the outage probability of the associated wireless-transmission links is also evaluated. The main contributions of this paper are boldly contrasted to the literature in Table I for visualizing the knowledge-gaps. Overall, in contrast to \cite{BIB04,BIB11,BIB13,BIB16,BIB17,BIB19,BIB20}, the effects of both the inter-constellation CFI and of the intra-constellation CFI are analyzed in this paper. To mitigate the effect of the inter-constellation CFI, multi-domain resources - including both the beam-domain and power-domain - are used for meeting the stringent requirements of a spectrum-sharing SGIN. We characterized them with the aid of their coverage analysis and their predicted CSI. Moreover, their closed-form outage probability (OP) and asymptotic OP (AOP) expressions were derived to assist in evaluating the overall performance of our proposed scheme. The main contributions of this paper are summarized as follows.

\begin{table*}
\label{tab:1}
\centering
\scriptsize
\caption{Boldly contrasting our novelty to the existing literature}
\begin{tabular}{l|c|c|c|c|c|c|c|c|c}
\bottomrule
&\cite{BIB04}&\cite{BIB11}&\cite{BIB13}&\cite{BIB16}&\cite{BIB17}&\cite{BIB19}&\cite{BIB20}&\textbf{Our Work}\\
\hline
Spectrum sharing &\checkmark& \checkmark &\checkmark  &\checkmark &\checkmark &\checkmark &\checkmark &$\pmb{\checkmark}$ \\
\hline
Among satellites (e.g., GEO and LEO, LEO and LEO) & \checkmark& \checkmark &  &\checkmark & & &\checkmark &$\pmb{\checkmark}$ \\
\hline
Satellites in coexistence with terrestrial systems &  &  & \checkmark &\checkmark & \checkmark&\checkmark & &$\pmb{\checkmark}$ \\
\hline
SGIN including GEO, LEO and terrestrial systems & &  &  & \checkmark& & & &$\pmb{\checkmark}$ \\
\hline
Coverage analysis &  &  & & & & & &$\pmb{\checkmark}$ \\
\hline
Power control & &  & \checkmark & & & \checkmark & \checkmark&$\pmb{\checkmark}$\\
\hline
Beam control &\checkmark &  &  & & \checkmark& &\checkmark &$\pmb{\checkmark}$ \\
\hline
Deep learning and CSI prediction aided resource scheduling & &  & & & & & &$\pmb{\checkmark}$ \\
\hline
Outage probability analysis & &  & \checkmark &\checkmark & & & &$\pmb{\checkmark}$ \\
\hline
Asymptotic outage probability analysis &  &  & & & & & &$\pmb{\checkmark}$ \\
\toprule
\end{tabular}
\end{table*}

{\textbf{Firstly}}, we propose joint multi-domain resource aided interference management (JMDR-IM) for a SGIN, relying on analyzing the coverage overlap, and on beneficially scheduling both the beam-domain and power-domain resources. Specifically, the coverage overlap is analyzed for identifying the potentially interfering satellites, and the alternative satellites available for substituting these interfering satellites. The equations derived may be beneficially used for designing a variety of similar systems. Moreover, the concept of co-frequency exclusion zone (CFEZ) is conceived.

{\textbf{Secondly}}, the activation of the beam-domain and power-domain resources is harmonized for managing the CFI by relying on our coverage overlap expressions. Specifically, the beam-domain scheduling based on the shutting off and switching beams is adopted for eliminating the CFI. Following the beam-domain scheduling issues, we conceive a long short-term memory based joint autoregressive moving average and deep Q network (LSTM-ARMA-DQN) aided power allocation scheme. Explicitly, this LSTM-ARMA-DQN scheme combines the LSTM-ARMA and the DQN philosophies, where the LSTM-ARMA is used for predicting the CSI of the satellite links, while the DQN is harnessed for optimizing the transmit power of our SGIN.

{\textbf{Thirdly}}, we present the OP and the AOP analysis of the proposed JMDR-IM scheme. Closed-form expressions are derived for the SGIN considered. Moreover, the asymptotic analysis of the OP and the AOP are also provided for further evaluating the performance of the proposed JMDR-IM scheme, and for verifying the accuracy of the OP and the AOP expressions derived.

{\textbf{Finally}}, we show that the proposed JMDR-IM scheme achieves a higher throughput despite its lower OP than that of the conventional pure stand-alone power allocation using fixed beams (PPAFB). Furthermore, the proposed JMDR-IM scheme maintains a high signal-to-interference plus noise ratio (SINR), which is achieved through tracking the dynamically fluctuating CFI of a SGIN.

The remainder of the paper is as follows. In Section II, we briefly present our system model, coverage analysis and the co-frequency exclusion zone concept. In Section III, we propose the JMDR-IM scheme, including our beam shut-off and beam-switching based scheduling, as well as the LSTM-ARMA-DQN based power allocation. In Section IV, we derive the OP of the proposed JMDR-IM scheme, while in Section V we conclude. The main variables and symbols used in this paper is listed in Table II for easy reference.

\begin{table}[htbp]
\scriptsize
\renewcommand\arraystretch{1}
\centering
\caption{Variable and Symbols List}
\begin{tabular}{p{0.2\textwidth}|p{0.25\textwidth}}
\bottomrule
Notation&Definition\\
\hline
NGSO 1, NGSO 2 & NGSO constellations\\
\hline
${\rm{N1}}_n$, ${\rm{N2}}_m$ & NGSO satellites\\
\hline
${\rm{BS}}_{i}$, ${\rm{BS}}_{z}$& Terrestrial base station (BS)\\
\hline
$p({{\rm{N1}}_n})$, $p({{\rm{N2}}_m})$ and $p({{\rm{BS}}_i})$& Locations of the NGSO satellites, and of the terrestrial BS\\
\hline
$p({{\rm{U}}_{\rm{N1}}^k})$, $p({{\rm{U}}_{\rm{N2}}^j})$, and $p({{\rm{U}}_{\rm{BS}}^t})$& Locations of users for the NGSO satellites, and of users for the terrestrial BS\\
\hline
${{\rm{U}}_{\rm{N1}}^k}$, ${{\rm{U}}_{\rm{N2}}^j}$ and ${{\rm{U}}_{\rm{BS}}^t}$& Users of satellites and BSs\\
\hline
$\{ {l_{{\rm{N1}}_n}^a,l_{{\rm{N1}}_n}^o} \}$, $\{ {l_{{\rm{U}}_{\rm{N1}}^k}^a,l_{{\rm{U}}_{\rm{N1}}^k}^o} \}$, \\$\{ {l_{{\rm{N2}}_m}^a,l_{{\rm{N2}}_m}^o} \}$, $\{ {l_{{\rm{U}}_{\rm{N2}}^j}^a,l_{{\rm{U}}_{\rm{N2}}^j}^o} \}$, \\$\{ {l_{{\rm{BS}}_{i}}^a,l_{{\rm{BS}}_{i}}^o} \}$, $\{ {l_{{\rm{U}}_{\rm{BS}}^t}^a,l_{{\rm{U}}_{\rm{BS}}^t}^o} \}$& Latitude and longitude\\
\hline
$P_n$, $P_m$, $P_i$, $P_z$, \\$P_n^o$, $P_m^o$, $P_i^o$, $P_z^o$& Transmit power\\
\hline
$h_{nk}$, $h_{mk}$, $h_{ik}$, $h_{mj}$, $h_{nj}$, $h_{ij}$\\$h_{it}$, $h_{zt}$, $h_{nt}$, $h_{mt}$ & Instantaneous CSI\\
\hline
$L_{nk}^S$, $L_{mk}^S$, $L_{ik}^S$, $L_{mj}^S$, $L_{nj}^S$,\\ $L_{ij}^S$, $L_{it}^S$, $L_{zt}^S$, $L_{nj}^S$, $L_{mj}^S$& Free-space loss\\
\hline
$G_{n{{\rm{U}}_{\rm{N1}}^k}}^t$, $G_{n{{\rm{U}}_{\rm{N1}}^k}}^r$, $G_{m{{\rm{U}}_{\rm{N1}}^k}}^t$, \\$G_{m{{\rm{U}}_{\rm{N1}}^k}}^r$, $G_{m{{\rm{U}}_{\rm{N2}}^j}}^t$, $G_{m{{\rm{U}}_{\rm{N2}}^j}}^r$,\\ $G_{n{{\rm{U}}_{\rm{N2}}^j}}^t$, $G_{n{{\rm{U}}_{\rm{N2}}^j}}^r$, $G_{n{{\rm{U}}_{\rm{BS}}^t}}^t$,\\ $G_{m{{\rm{U}}_{\rm{BS}}^t}}^t$& Antenna gain\\
\hline
$x_{\rm{N1}}^{n}$, $x_{\rm{N2}}^{m}$, $x_{\rm{BS}}^{i}$, $x_{\rm{BS}}^{z}$ & Transmitted signals\\
\hline
${\alpha _m}$, ${\alpha _n}$ & Traffic states\\
\hline
$n_k$, $n_j$, $n_t$ & Gaussian noise\\
\hline
$N_{{\rm{S1}}}$, $N_{{\rm{S2}}}$, $N_{{\rm{S3}}}$, $N_{{\rm{S4}}}$, \\$N_{{\rm{S1}}}^{'}$, $N_{{\rm{S2}}}^{'}$, $N_{{\rm{S3}}}^{'}$, $N_{{\rm{S4}}}^{'}$& The number of interfering satellites \\
\hline
$b_{{\rm{N1}}_n}^l$& Satellites' beams\\
\hline
${\Phi _{{{\rm{N2}}_m} \to {{\rm{U}}_{\rm{N1}}^k}}}$, ${\Phi _{{{\rm{N2}}_m} \to {{\rm{U}}_{\rm{N2}}^j}}}$, \\${\Phi _{{{\rm{N2}}_m} \to {{\rm{U}}_{\rm{BS}}^t}}}$, ${\Phi _{{{\rm{N1}}_n} \to {{\rm{U}}_{\rm{N1}}^k}}}$, \\${\Phi _{{{\rm{N1}}_n} \to {{\rm{U}}_{\rm{N2}}^j}}}$, ${\Phi _{{{\rm{N1}}_n} \to {{\rm{U}}_{\rm{BS}}^t}}}$& Satellites' sets of covering users\\
\hline
$P_{\rm{NGSO1}}^{\max }$, $P_{\rm{NGSO2}}^{\max }$, $P_{\rm{BS}}^{\max }$ & Maximum available transmit power\\
\hline
LEO, GEO & Low-Earth-Orbit, Geostationary\\
\hline
NGSO & Nongeostationary\\
\hline
SGIN & Satellite-ground integrated network\\
\hline
CFI, CSI & Co-frequency interference, Channel state information\\
\hline
CFEZ & Co-frequency exclusion zone\\
\hline
JMDR-IM & Joint multi-domain resource aided interference management\\
\hline
LSTM-ARMA-DQN & Long short-term memory based joint autoregressive moving average and deep Q network\\
\hline
PPAFB & Pure stand-alone power allocation using fixed beams\\
\hline
OP, AOP & Outage probability, Asymptotic outage probability\\
\hline
SINR & Signal-to-interference plus noise ratio\\
\toprule
\end{tabular}
\end{table}
\hfill

\section{System Model and Coverage Analysis}

\subsection{System Model}
As shown in Fig.~\ref{fig:fig1}, we investigate a spectrum-sharing SGIN, composed of two LEO constellations, represented by ${\rm{N1}}_n$ and ${\rm{N2}}_m$, and multiple terrestrial BSs, denoted by ${\rm{BS}}_{i}$, where $1 \le n \le N$, $1 \le m \le M$, $1 \le i \le N_B$. Explicitly, $N$, $M$ and $N_B$ are the number of NGSO 1 satellites, of NGSO 2 satellites and of the BSs, respectively. Moreover, their users are denoted by ${{\rm{U}}_{\rm{N1}}^k}$, ${{\rm{U}}_{\rm{N2}}^j}$ and ${{\rm{U}}_{\rm{BS}}^t}$, where $1 \le k \le K$, $1 \le j \le J$ and $1 \le t \le N_t$. Furthermore, $K$, $J$ and $N_t$ are the number of users of the NGSO 1, of the NGSO 2 and of the BSs, respectively. The location of each satellite is represented by the latitude and the longitude. Thus, given the $n$-th satellite of NGSO 1 and the $k$-th user, their locations are $p( {\rm{N1}}_n) = \{ {l_{{\rm{N1}}_n}^a,l_{{\rm{N1}}_n}^o} \}$ and $p( {{\rm{U}}_{\rm{N1}}^k}) = \{ {l_{{\rm{U}}_{\rm{N1}}^k}^a,l_{{\rm{U}}_{\rm{N1}}^k}^o} \}$. Similarly, given the $m$-th satellite of NGSO 2 and the $j$-th user, their locations are $p( {\rm{N2}}_m) = \{ {l_{{\rm{N2}}_m}^a,l_{{\rm{N2}}_m}^o} \}$ and $p( {{\rm{U}}_{\rm{N2}}^j}) = \{ {l_{{\rm{U}}_{\rm{N2}}^j}^a,l_{{\rm{U}}_{\rm{N2}}^j}^o} \}$. Let $p( {\rm{BS}}_{i}) = \{ {l_{{\rm{BS}}_{i}}^a,l_{{\rm{BS}}_{i}}^o} \}$ and $p( {{\rm{U}}_{\rm{BS}}^t}) = \{ {l_{{\rm{U}}_{\rm{BS}}^t}^a,l_{{\rm{U}}_{\rm{BS}}^t}^o} \}$ denote the locations of the $i$-th BS and its $t$-th user, respectively. We also assume that the channel spanning from any satellite to any user is accurately represented by a Shadowed-Rician fading model \cite{BIB23}. Moreover, we use a Rayleigh fading channel for modeling all the links between any BS and any user. We also use a Poisson model for characterizing the tele-traffic of both NGSO 1 and NGSO 2 \cite{BIBPoisson}. Furthermore, we consider a scenario, where NGSO 1 and NGSO 2 occupy two different orbits, and they are equipped with switchable multi-beam antennas capable of pointing in specific directions \cite{BIB24}.

\begin{figure}[htbp]
\centering
\includegraphics[width=0.9\linewidth]{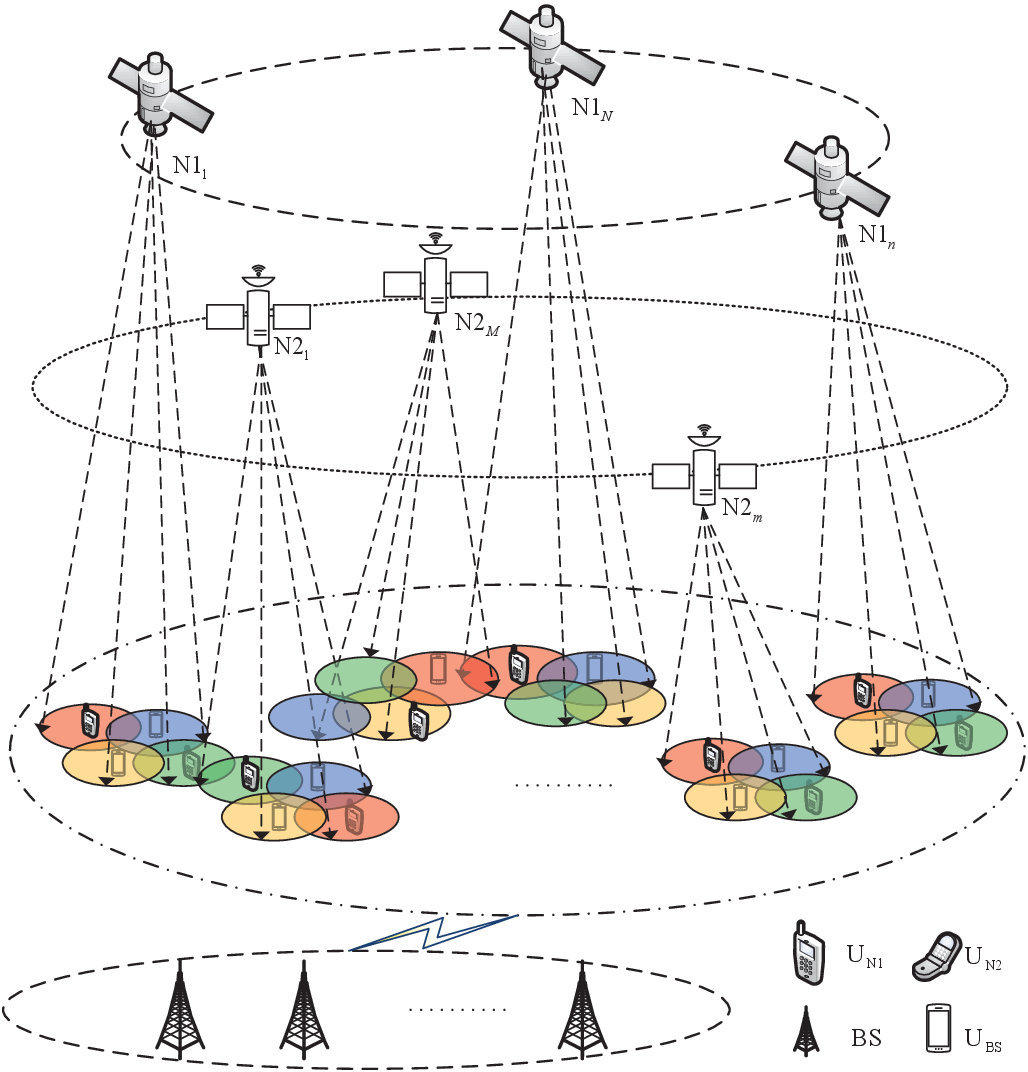}
\caption{A spectrum-sharing SGIN, including a pair of NGSO constellations and multiple terrestrial BSs.\label{fig:fig1}}
\end{figure}

As shown in Fig.~\ref{fig:fig1}, we assume that the multi-color multibeam frequency-reuse scheme is employed by both NGSO 1 and NGSO 2. For notational convenience, we consider single-color beams for representing all multi-color beams as an example, but all remaining beams can be expressed similarly. We assume that given the $n$-th satellite, a beam of this satellite transmits its signal to the $k$-th user of NGSO 1, thus the signal received at ${{\rm{U}}_{\rm{N1}}^k}$ is given by
\begin{eqnarray}
&&\!\!\!\!\!\!\!\!\!\!\!\!{y_{{\rm{U}}_{\rm{N1}}^k}} \!=\! \sqrt {{P_{n}}G_{n{{\rm{U}}_{\rm{N1}}^k}}^tG_{n{{\rm{U}}_{\rm{N1}}^k}}^rL_{nk}^S} {h_{nk}}{x_{\rm{N1}}^{n}} \!\!+ \!\!\sum\limits_{i \in \Phi _{\rm{BS}}}\!\! {\sqrt {{P_{i}}L_{ik}^{S}} {h_{ik}}{x_{\rm{BS}}^{i}}} \nonumber\\
&&+\!\!\sum\limits_{m = 1}^{{N_{\rm{S1}}}}\!\! {\sqrt {{\alpha _m}{P_{m}}G_{m{{\rm{U}}_{\rm{N1}}^k}}^tG_{m{{\rm{U}}_{\rm{N1}}^k}}^rL_{mk}^S} {h_{mk}}{x_{\rm{N2}}^{m}}}
   + {n_k}, \label{equ:1}
\end{eqnarray}
where $P_n$, $P_m$ and $P_i$ denote the transmit power of ${\rm{N1}}_n$, ${\rm{N2}}_m$ and ${\rm{BS}}_i$, respectively. Furthermore, $x_{\rm{N1}}^{n}$, $x_{\rm{N2}}^{m}$ and $x_{\rm{BS}}^{i}$ respectively denote the random symbols transmitted by the $n$-th satellite of NGSO 1, the $m$-th satellite of NGSO 2 and the $i$-th BS of the BS set. We also assume $E[|x_{\rm{N1}}^{n}|^2] = E[|x_{\rm{N2}}^{m}|^2] = E[|x_{\rm{BS}}^{i}|^2] = 1$, where $E[\cdot]$ represents the expected value. The variable $N_{\rm{S1}}$ represents the number of satellites imposing interference on ${{\rm{U}}_{\rm{N1}}^k}$, while $n_k$ is the zero-mean noise received at ${{\rm{U}}_{\rm{N1}}^k}$ having a variance of $k_BB_nT_n$, with $k_B$, $B_n$ and $T_n$ denoting the Boltzmann constant, the transmission bandwidth of ${\rm{N1}}_n$'s beam and the equivalent noise temperature, respectively. The variables $h_{nk}$, $h_{mk}$ and $h_{ik}$ denote the instantaneous CSI of the links spanning from ${\rm{N1}}_n$ to ${{\rm{U}}_{\rm{N1}}^k}$, from ${\rm{N2}}_m$ to ${{\rm{U}}_{\rm{N1}}^k}$ and from ${\rm{BS}}_i$ to ${{\rm{U}}_{\rm{N1}}^k}$, respectively. Additionally, $L_{nk}^S$, $L_{mk}^S$ and $L_{ik}^S$ represent the free-space loss of the ${\rm{N1}}_n$-${{\rm{U}}_{\rm{N1}}^k}$, ${\rm{N2}}_m$-${{\rm{U}}_{\rm{N1}}^k}$ and ${\rm{BS}}_i$-${{\rm{U}}_{\rm{N1}}^k}$ links \cite{BIBCOVER}, while ${\alpha _m}$ represents the traffic state of ${\rm{N2}}_m$'s beam, where the traffic state is assumed to obey the Poisson distribution \cite{BIBPoisson}. If there is no tele-traffic during a given transmission slot, we have ${\alpha _m} = 0$. Otherwise, we have ${\alpha _m} = 1$. Furthermore, $G_{ab}^d$ denotes the antenna gain of the satellites and their users, and we have $G_{ab}^d = {G_0}{[\frac{{{J_1}(\mu )}}{{2\mu }} + 36\frac{{{J_3}(\mu )}}{{{\mu ^3}}}]^2}$ \cite{BIB20}, where ${{\mu  = 2.07123\sin (\theta )} \mathord{\left/
 {\vphantom {{\mu  = 2.07123\sin (\theta )} {\sin ({\theta _{\rm{3 dB}}})}}} \right.
 \kern-\nulldelimiterspace} {\sin ({\theta _{\rm{3 dB}}})}}$, $\theta$ and $\theta _{\rm{3 dB}}$ respectively represent the off-boresight angle and the 3 dB angle. In this antenna-gain expression, $J_1$ and $J_3$ denote the Bessel functions of the first order and the third order, respectively, while ${G_0} = \xi {(\frac{{\pi {D_e}}}{{c \mathord{\left/
 {\vphantom {c f}} \right.
 \kern-\nulldelimiterspace} f} })^2}$, $D_e$, $c$, $f$ and $\xi$ respectively denote the antenna diameter, the light velocity, the frequency band and the antenna efficiency. Finally, $a\in \{n,m\}$, $b\in \{{{\rm{U}}_{\rm{N2}}^j},{{\rm{U}}_{\rm{N1}}^k}\}$, $d\in \{t,r\}$ and $e\in \{{\rm{N1}}_n,{\rm{N2}}_m,{{\rm{U}}_{{\rm{N1}}_k}},{{\rm{U}}_{{\rm{N2}}_j}}\}$.

Similarly, given the $m$-th NGSO 2 satellite and the $j$-th NGSO 2 user, the signal received at ${{\rm{U}}_{\rm{N2}}^j}$ is formulated as
\begin{eqnarray}
&&\!\!\!\!\!\!\!\!\!\!\!\!{y_{{\rm{U}}_{\rm{N2}}^j}} \!=\! \sqrt {{P_{m}}\!G_{m{{\rm{U}}_{\rm{N2}}^j}}^t\!G_{m{{\rm{U}}_{\rm{N2}}^j}}^r\!\!L_{mj}^S} {h_{mj}}{x_{\rm{N2}}^{m}} \!\!+\!\!\! \sum\limits_{i \in \Phi _{\rm{BS}}} \!\!{\sqrt {{P_{i}}L_{ij}^{S}} {h_{ij}}{x_{\rm{BS}}^{i}}}\nonumber\\
&&+\!\sum\limits_{n = 1}^{{N_{{\rm{S2}}}}} \!\!{\sqrt {{\alpha _n}{P_{n}}G_{n{{\rm{U}}_{\rm{N2}}^j}}^tG_{n{{\rm{U}}_{\rm{N2}}^j}}^rL_{nj}^S} {h_{nj}}{x_{\rm{N1}}^{n}}}+ {n_j}, \label{equ:2}
\end{eqnarray}
where $N_{{\rm{S2}}}$ is the number of satellites imposing interference on ${{\rm{U}}_{\rm{N2}}^j}$. In (2), $n_j$ is the noise received at ${{\rm{U}}_{\rm{N2}}^j}$ having a zero mean and a variance of $k_BB_mT_n$, $B_m$ is the transmission bandwidth of ${\rm{N2}}_m$'s beam. Furthermore, $h_{mj}$, $h_{nj}$ and $h_{ij}$ represent the CSIs of the links from ${\rm{N2}}_m$ to ${{\rm{U}}_{\rm{N2}}^j}$, from ${\rm{N1}}_n$ to ${{\rm{U}}_{\rm{N2}}^j}$ and from ${\rm{BS}}_i$ to ${{\rm{U}}_{\rm{N2}}^j}$, respectively. Still referring to (2), $L_{mj}^S$, $L_{nj}^S$ and $L_{ij}^S$ represent the free-space loss of the ${\rm{N2}}_m$-${{\rm{U}}_{\rm{N2}}^j}$, ${\rm{N1}}_n$-${{\rm{U}}_{\rm{N2}}^j}$ and ${\rm{BS}}_i$-${{\rm{U}}_{\rm{N2}}^j}$ links. Similarly to ${\alpha _m}$, ${\alpha _n}$ denotes the traffic state of ${\rm{N1}}_n$'s beam, and we have ${\alpha _n} \in \{0,1\}$.

To elaborate further, the signal received at ${{\rm{U}}_{\rm{BS}}^t}$ and transmitted by ${\rm{BS}}_i$ is written as
\begin{eqnarray}
&&\!\!\!\!\!\!\!\!\!\!\!\!\!\!\!\!\!\!\!\!{y_{{\rm{U}}_{\rm{BS}}^t}} \!=\! \sqrt {{P_{i}}L_{it}^S} {h_{it}}{x_{\rm{BS}}^{i}} +\!
\sum\limits_{n = 1}^{{N_{{\rm{S3}}}}} \!{\sqrt {{\alpha _n}{P_{n}}G_{n{{\rm{U}}_{\rm{BS}}^t}}^tL_{nt}^S} {h_{nt}}{x_{\rm{N1}}^{n}}} \nonumber\\
&&\!\!\!\!\!+\sum\limits_{m = 1}^{{N_{{\rm{S4}}}}} \!\!{\sqrt {{\alpha _m}\!{P_{m}}G_{m{{\rm{U}}_{\rm{BS}}^t}}^t\!L_{mt}^S} {h_{mt}}{x_{\rm{N2}}^{m}}}\nonumber \\
&&\!\!\!\!\!+\sum\limits_{z \in \Phi _{\rm{BS}},z \ne i} {\sqrt {{P_{z}}\!L_{zt}^{S}} {h_{zt}}{x_{\rm{BS}}^{z}}} +{n_t}, \label{equ:3}
\end{eqnarray}
where $N_{{\rm{S3}}}$ and $N_{{\rm{S4}}}$ respectively denote the number of NGSO 1 satellites and NGSO 2 satellites imposing interference on ${{\rm{U}}_{\rm{BS}}^t}$. Furthermore, $n_t$ is the noise received at ${{\rm{U}}_{\rm{BS}}^t}$ having a zero mean and a variance of $\sigma _n^2$. In (3), $h_{it}$, $h_{zt}$, $h_{nt}$ and $h_{mt}$ represent the CSIs of the links spanning from ${\rm{BS}}_i$ to ${{\rm{U}}_{\rm{BS}}^t}$, from ${\rm{BS}}_z$ to ${{\rm{U}}_{\rm{BS}}^t}$, from ${\rm{N1}}_n$ to ${{\rm{U}}_{\rm{BS}}^t}$ and from ${\rm{N2}}_m$ to ${{\rm{U}}_{\rm{BS}}^t}$, respectively. Moreover, $L_{it}^S$, $L_{zt}^S$, $L_{nj}^S$, and $L_{mj}^S$ represent the free-space loss of the ${\rm{BS}}_i$-${{\rm{U}}_{\rm{BS}}^t}$, ${\rm{BS}}_z$-${{\rm{U}}_{\rm{BS}}^t}$, ${\rm{N1}}_n$-${{\rm{U}}_{\rm{BS}}^t}$ and ${\rm{N2}}_m$-${{\rm{U}}_{\rm{BS}}^t}$ links, respectively. Moreover, in (3), the inter-cell CFI is considered \cite{BIBICI1,BIBICI2}.

Thus, using (1), (2) and (3), we can express the SINR of the ${\rm{N1}}_n$-${{\rm{U}}_{\rm{N1}}^k}$, ${\rm{N2}}_m$-${{\rm{U}}_{\rm{N2}}^j}$ and ${\rm{BS}}_i$-${{\rm{U}}_{\rm{BS}}^t}$ links as
\begin{eqnarray}
&&\!\!\!\!\!\!\!\!\!\!\!\!\!\!\!\!\!\!{\rm{SINR}}_{{\rm{U}}_{\rm{N1}}^k} =\nonumber\\ &&\!\!\!\!\!\!\!\!\!\!\!\!\!\!\!\!\!\!\frac{{{P_{n}}G_{n{{\rm{U}}_{\rm{N1}}^k}}^tG_{n{{\rm{U}}_{\rm{N1}}^k}}^rL_{nk}^S}{{\left| {{h_{nk}}} \right|}^2}}{{\sum\limits_{m = 1}^{{N_{\rm{S1}}}}\!\!{{\alpha _m}\!{P_{m}}\!G_{\!m\!{{\rm{U}}_{\rm{N1}}^k}}^t\!\!G_{\!\!m\!{{\rm{U}}_{\rm{N1}}^k}}^r\!\!L_{mk}^S}{{\left|\! {{h_{mk}}} \!\right|}^2} \!\!+\!\!\!\! \sum\limits_{i \in \Phi _{\rm{BS}}}\!\!\!{{P_{i}}L_{ik}^S}{{\left| {{h_{ik}}} \right|}^2} \!\!+\! k_B\!B_n\!T_n}} \label{equ:4}
\end{eqnarray}
and
\begin{eqnarray}
&&\!\!\!\!\!\!\!\!\!\!\!\!\!\!\!\!\!\!{\rm{SINR}}_{{\rm{U}}_{\rm{N2}}^j} = \nonumber\\ &&\!\!\!\!\!\!\!\!\!\!\!\!\!\!\!\!\!\!\frac{{{P_{m}}G_{m{{\rm{U}}_{\rm{N2}}^j}}^tG_{m{{\rm{U}}_{\rm{N2}}^j}}^rL_{mj}^S}{{\left| {{h_{mj}}} \right|}^2}}{{\sum\limits_{n = 1}^{{N_{{\rm{S2}}}}}\!\!{{\alpha _n}\!{P_{n}}\!G_{\!n\!{{\rm{U}}_{\rm{N2}}^j}}^t\!G_{\!n\!{{\rm{U}}_{\rm{N2}}^j}}^r\!\!L_{nj}^S}{{\left|\! {{h_{nj}}} \!\right|}^2} \!\!+\!\!\!\! \sum\limits_{i \in \Phi _{\rm{BS}}}\!\!\!{{P_{i}}\!L_{ij}^S}\!{{\left| {{h_{ij}}} \right|}^2} \!\!+\! k_B\!B_m\!T_n}} \label{equ:5}
\end{eqnarray}
and
\begin{eqnarray}
&&\!\!\!\!\!\!\!\!\!\!\!\!\!\!\!\!\!\!\!\!\!{\rm{SINR}}_{{\rm{U}}_{\rm{BS}}^t} = \nonumber\\
&&\!\!\!\!\!\!\!\!\!\!\!\!\!\!\!\!\!\!\!\!\!\frac{{{P_{i}}L_{it}^S}{{\left| {{h_{it}}} \right|}^2}}{{\sum\limits_{n = 1}^{{N_{{\rm{S3}}}}}\!\!{{\alpha _n}\!{P_{n}}\!L_{nt}^S}{{\left|\! {{h_{nt}}} \!\right|}^2} \!\!+\!\!\! \sum\limits_{m = 1}^{{N_{{\rm{S4}}}}}\!\!\!{{\alpha _m}\!{P_{m}}\!L_{mt}^S}\!{{\left| {{h_{mt}}} \right|}^2}\!\!\!+\!\!\!\!\!\!\sum\limits_{z \!\in\! \Phi _{\rm{BS}}, z \!\ne\! i}\!\!\!\!\!{{P_{z}}\!L_{zt}^S}\!{{\left|\! {{h_{zt}}} \!\right|}^2} \!\!+\!\! \sigma _n^2}}.\label{equ:6}
\end{eqnarray}

Generally, although the spectrum sharing between NGSO 1 and NGSO 2 is capable of improving the spectrum efficiency, the coverage overlap between NGSO 1 and NGSO 2 may impose severe interference. As shown in Fig.~\ref{fig:fig2}, this interference is analyzed for a pair of NGSO constellations, wherein $C/N$ and $C/(I+N)$ denote the carrier to noise ratio, and carrier-to-interference plus noise ratio, respectively. Observe from Fig.~\ref{fig:fig2} that NGSO 2 may impose excessive interference on NGSO 1, which fluctuates as these satellites move.
\begin{figure}[htbp]
\centering
\includegraphics[scale=0.6]{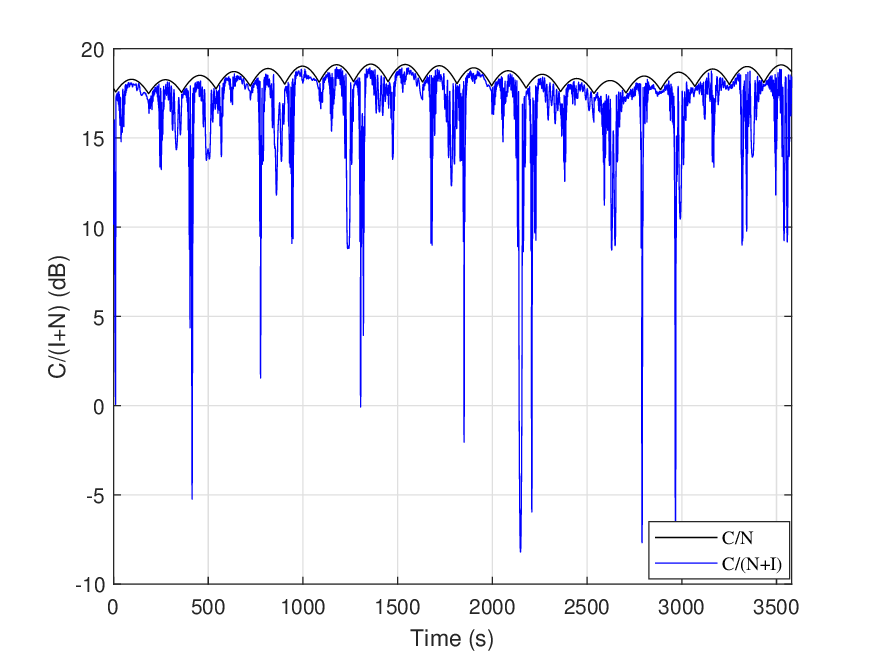}
\caption{Co-frequency interference at a typical NGSO 1 imposed by a spectrum-sharing NGSO 2, where the number of satellites in NGSO 1 and NGSO 2 are 648 and 1584, respectively.\label{fig:fig2}}
\end{figure}

As shown in Fig.~\ref{fig:fig3}, the intra-constellation CFI of both NGSO 1 and NGSO 2 exist, when the four-color frequency-reuse (FR4) multibeam scheme \cite{BIBFR4} is adopted. By contrast, the intra-constellation CFI of both NGSO 1 and NGSO 2 becomes negligible in the presence of the seven-color frequency-reuse (FR7) multibeam scheme \cite{BIBFR7} as shown in Fig.~\ref{fig:fig3}. This beneficial result is achieved, since the reuse protection distance of the same frequency is high enough. We should point out that the inter-constellation CFI is neglected in Fig.~\ref{fig:fig3}, and we only take the intra-constellation CFI into account. By contrast, the inter-constellation CFI may be more severe than the intra-constellation CFI. Thus, we focus our attention on the inter-constellation CFI in this paper.

\begin{figure}[htbp]
\centering
\includegraphics[scale=0.6]{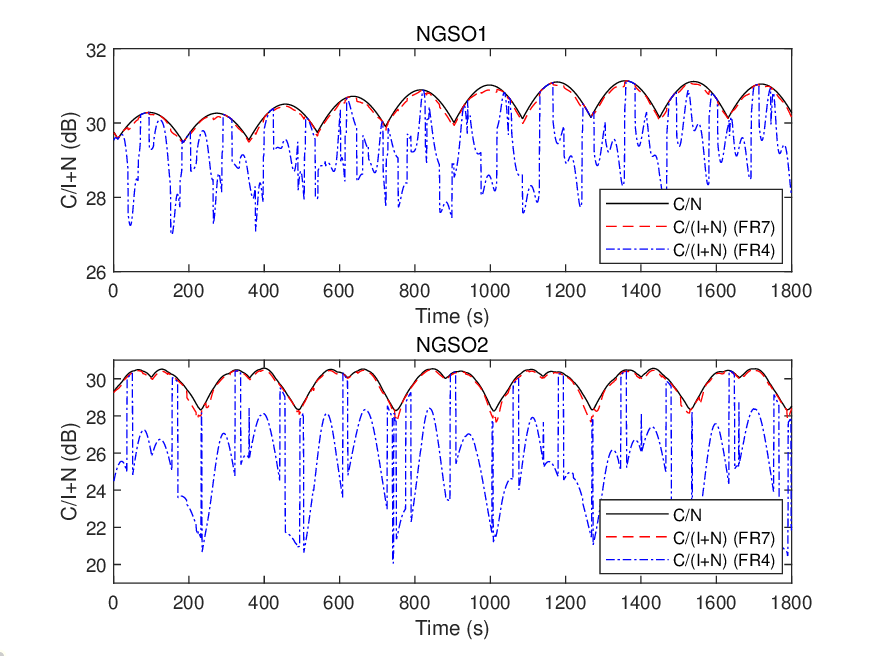}
\caption{Intra-constellation co-frequency interference at a typical NGSO 1 and NGSO 2, wherein the FR4 multibeam and the FR7 multibeam are respectively analyzed, the number of satellites in NGSO 1 and NGSO 2 are respectively 648 and 1584, and the inter-constellation co-frequency interference is neglected.\label{fig:fig3}}
\end{figure}

\subsection{Coverage Analysis}

In this subsection, we analyze the satellite coverage. Given the $l$-th beam of the satellite ${\rm{N1}}_n$, denoted by $b_{{\rm{N1}}_n}^l$, its coverage is shown in Fig.~\ref{fig:fig4}.

\begin{figure}[htbp]
\centering
\includegraphics[scale=0.5]{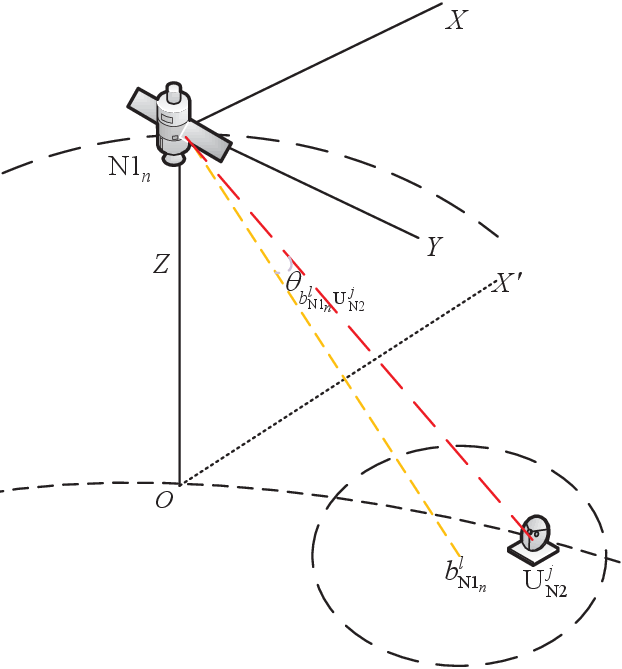}
\caption{The coverage analysis for an NGSO 1 satellite.\label{fig:fig4}}
\end{figure}

Given ${{\rm{N1}}_n}$ and ${{\rm{U}}_{\rm{N2}}^j}$, the condition of ${{\rm{N1}}_n}$ covering ${{\rm{U}}_{\rm{N2}}^j}$ can be summarized as
\begin{eqnarray}
&&\!\!\!\!\!\!\!\!\!\!\!\!\!\!\!\!\!\!\!\!\!\!\!\!\!\!{C_{{{\rm{N1}}_n}{{\rm{U}}_{\rm{N2}}^j}}} =\left\{ {\left( {l_{{{\rm{U}}_{\rm{N2}}^j}}^a,l_{{{\rm{U}}_{\rm{N2}}^j}}^o,l_{{\rm{N1}}_n}^o} \right)}:\right.\nonumber \\
&&\!\!\!\!\!\!\!\!\!\!{\theta ^E}\left( {l_{{{\rm{U}}_{\rm{N2}}^j}}^a,l_{{{\rm{U}}_{\rm{N2}}^j}}^o,l_{{\rm{N1}}_n}^o} \right) > {0^ \circ } \nonumber\\
&&\!\!\!\!\!\!\!\!\!\!\&{\theta ^A}\left( {l_{{{\rm{U}}_{\rm{N2}}^j}}^a,l_{{{\rm{U}}_{\rm{N2}}^j}}^o,l_{{\rm{N1}}_n}^o} \right) \in \left[ { - {{180}^ \circ },{{180}^ \circ }} \right]\nonumber\\
&&\!\!\!\!\!\!\!\!\!\! \&{d_{{{\rm{N1}}_n}{{\rm{U}}_{\rm{N2}}^j}}} \le d_{{\rm{N1}}_n}^{\max }{\rm{                 }}\left. {\&{\theta _{{{b_{{\rm{N1}}_n}^l}}{{\rm{U}}_{\rm{N2}}^j}}} \le \frac{{{\theta ^{\rm{BW}}}}}{2}} \right\},\label{equ:7}
\end{eqnarray}
where ${\theta ^E}( {l_{{{\rm{U}}_{\rm{N2}}^j}}^a,l_{{{\rm{U}}_{\rm{N2}}^j}}^o,l_{{\rm{N1}}_n}^o} )$ and ${\theta ^A}( {l_{{{\rm{U}}_{\rm{N2}}^j}}^a,l_{{{\rm{U}}_{\rm{N2}}^j}}^o,l_{{\rm{N1}}_n}^o} )$ denote the elevation angle and the azimuth angle from ${{\rm{U}}_{\rm{N2}}^j}$ to ${{\rm{N1}}_n}$, respectively. Furthermore, ${d_{{{\rm{N1}}_n}{{\rm{U}}_{\rm{N2}}^j}}}$ is the distance between ${{\rm{N1}}_n}$ and ${{\rm{U}}_{\rm{N2}}^j}$, while $d_{{\rm{N1}}_n}^{\max }$ represents the distance between ${{\rm{N1}}_n}$ and the border of its beam ${b_{{\rm{N1}}_n}^l}$. Still referring to the above equation, ${\theta^{\rm{BW}}}$ is the covering angle of a beam of the ${\rm{N1}}_n$, and ${\theta _{{b_{{\rm{N1}}_n}^l}{{\rm{U}}_{\rm{N2}}^j}}}$ denotes the angle between the link ${{\rm{N1}}_n} \to$ the center of ${b_{{\rm{N1}}_n}^l}$ and the link ${\rm{N1}}_n \to$ the ${{\rm{U}}_{\rm{N2}}^j}$. Similarly, given ${{\rm{N1}}_n}$, ${{\rm{N2}}_m}$, ${{\rm{U}}_{\rm{N1}}^k}$, and ${{\rm{U}}_{\rm{BS}}^t}$, the conditions of ${{\rm{N2}}_m}$ covering ${{\rm{U}}_{\rm{N1}}^k}$, of ${{\rm{N1}}_n}$ covering ${{\rm{U}}_{\rm{BS}}^t}$, of ${{\rm{N2}}_m}$ covering ${{\rm{U}}_{\rm{BS}}^t}$ can be summarized.

To elaborate further, given ${{\rm{U}}_{\rm{N1}}^k}$, ${{\rm{U}}_{\rm{N2}}^j}$ and ${{\rm{U}}_{\rm{BS}}^t}$, using the above coverage conditions, the NGSO 2 satellite set ${\Phi _{{{\rm{N2}}_m} \to {{\rm{U}}_{\rm{N1}}^k}}}$ covering ${{\rm{U}}_{\rm{N1}}^k}$, the NGSO 2 satellite set ${\Phi _{{{\rm{N2}}_m} \to {{\rm{U}}_{\rm{N2}}^j}}}$ covering ${{\rm{U}}_{\rm{N2}}^j}$, the NGSO 2 satellite set ${\Phi _{{{\rm{N2}}_m} \to {{\rm{U}}_{\rm{BS}}^t}}}$ covering ${{\rm{U}}_{\rm{BS}}^t}$, the NGSO 1 satellite set ${\Phi _{{{\rm{N1}}_n} \to {{\rm{U}}_{\rm{N1}}^k}}}$ covering ${{\rm{U}}_{\rm{N1}}^k}$, the NGSO 1 satellite set ${\Phi _{{{\rm{N1}}_n} \to {{\rm{U}}_{\rm{N2}}^j}}}$ covering ${{\rm{U}}_{\rm{N2}}^j}$ and the NGSO 1 satellite set ${\Phi _{{{\rm{N1}}_n} \to {{\rm{U}}_{\rm{BS}}^t}}}$ covering ${{\rm{U}}_{\rm{BS}}^t}$ can finally be obtained. By exploiting the number of satellites in the sets ${\Phi _{{{\rm{N2}}_m} \to {{\rm{U}}_{\rm{N1}}^k}}}$, ${\Phi _{{{\rm{N1}}_n} \to {{\rm{U}}_{\rm{N2}}^j}}}$, ${\Phi _{{{\rm{N1}}_n} \to {{\rm{U}}_{\rm{BS}}^t}}}$ and ${\Phi _{{{\rm{N2}}_m} \to {{\rm{U}}_{\rm{BS}}^t}}}$, $N_{\rm{S1}}$, $N_{\rm{S2}}$, $N_{\rm{S3}}$ and $N_{\rm{S4}}$ can also be found.

\subsection{Co-Frequency Exclusion Zone}

Based on the above coverage analysis, we exploit the CFEZ for CFI management, especially for managing the co-line interference. Specifically, a pair of spectrum-sharing satellites in a CFEZ may impose excessive interference on each others' users, which may result in call-dropping. As shown in Fig.~\ref{fig:fig5}, we consider the NGSO 1 satellite ${{\rm{N1}}_n}$, the NGSO 2 satellite ${{\rm{N2}}_m}$, and their users ${{\rm{U}}_{\rm{N1}}^k}$ and ${{\rm{U}}_{\rm{N2}}^j}$. Moreover, an angle $\theta _{{{{\rm{N1}}_n}}{{\rm{U}}_{\rm{N1}}^k}{{{\rm{N2}}_m}}}$, between ${{\rm{N1}}_n}\to$${{\rm{U}}_{\rm{N1}}^k}$ and ${{\rm{N2}}_m}\to$${{\rm{U}}_{\rm{N1}}^k}$, is used to detect the event that ${{\rm{N2}}_m}$ enters into the CFEZ of ${{\rm{N1}}_n}$. Such an angle $\theta _{{{{\rm{N1}}_n}}{{\rm{U}}_{\rm{N1}}^k}{{{\rm{N2}}_m}}}$ is formulated as
\begin{equation}
{\theta _{{{\rm{N1}}_n}{{\rm{U}}_{\rm{N1}}^k}{{\rm{N2}}_m}}} \!=\! \arccos\! \left(\! {\frac{{d_{{{\rm{N1}}_n}{{\rm{U}}_{\rm{N1}}^k}}^2 \!+ \! d_{{{\rm{N2}}_m}{{\rm{U}}_{\rm{N1}}^k}}^2 \!-\! d_{{{\rm{N1}}_n}{{\rm{N2}}_m}}^2}}{{2  {d_{{{\rm{N1}}_n}{{\rm{U}}_{\rm{N1}}^k}}}  {d_{{{\rm{N2}}_m}{{\rm{U}}_{\rm{N1}}^k}}}}}} \!\right), \label{equ:8}
\end{equation}
where $d_{{{\rm{N1}}_n}{{\rm{N2}}_m}}$ can be expressed by
\begin{eqnarray}
&&\!\!\!\!\!\!\!\!\!\!\!\!\!\!\!\!\!\!\!\!\!{d_{{{\rm{N1}}_n}\!{{\rm{N2}}_m}}} =\nonumber \\
&&\!\!\!\!\!\!\!\!\!\!\!\!\!\!\!\!\!\!\!\!\!\sqrt {{{\left(\! {{X_{{{\rm{N1}}_n}}} \!\!-\!\! {X_{{{\rm{N2}}_m}}}} \!\right)}^2} \!+\! {{\left(\! {{Y_{{{\rm{N1}}_n}}} \!\!-\!\! {Y_{{{\rm{N2}}_m}}}} \!\right)}^2} \!+\! {{\left(\! {{Z_{{{\rm{N1}}_n}}} \!\!-\!\! {Z_{{{\rm{N2}}_m}}}} \!\right)}^2}},  \label{equ:9}
\end{eqnarray}
where $\{{X_{{{\rm{N1}}_n}}},{Y_{{{\rm{N1}}_n}}},{Z_{{{\rm{N1}}_n}}}\}$ and $\{{X_{{{\rm{N2}}_m}}},{Y_{{{\rm{N2}}_m}}},{Z_{{{\rm{N2}}_m}}}\}$ respectively denote the locations of ${{\rm{N1}}_n}$ and ${{\rm{N2}}_m}$, which can be specified in the earth-centered earth-fixed coordinate system. Given the calculated $\theta _{{{{\rm{N1}}_n}}{{\rm{U}}_{\rm{N1}}^k}{{{\rm{N2}}_m}}}$, if we have $\theta _{{{{\rm{N1}}_n}}{{\rm{U}}_{\rm{N1}}^k}{{{\rm{N2}}_m}}}<\theta _{\rm{CFEZ}}$, then spectrum sharing between these two satellites is disabled, and this zone is defined as the CFEZ.

\begin{figure}[htbp]
\centering
\includegraphics[width=0.8\linewidth]{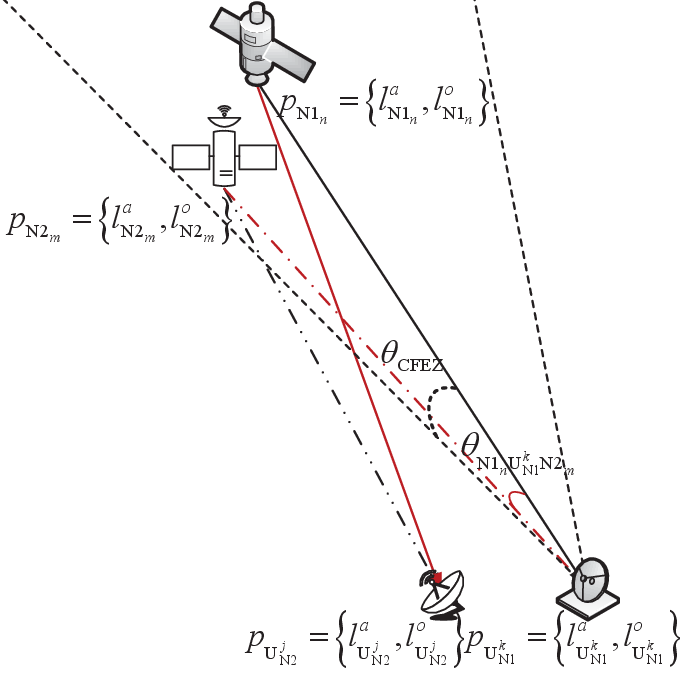}
\caption{Co-frequency exclusion zone analysis.\label{fig:fig5}}
\end{figure}

\subsection{Problem Formulation}

In order to support spectrum sharing for the SGIN considered while managing the corresponding CFI, the optimization problem is formulated as
\begin{eqnarray}
&&\mathop {\max }\limits_{{P_{n}},{P_{m}},{P_{i}},{N_{\rm{S1}}},{N_{\rm{S2}}},{N_{\rm{S3}}},{N_{\rm{S4}}}} {\rm{SINR}}_{{{\rm{U}}_{\rm{N1}}^k}}\nonumber \\
&&{\rm{         s.t.}}\;\;\;\;\;\;\;\;\;\;\;\;{\rm{         0}} \le {P_{n}} \le P_{\rm{NGSO1}}^{\max }\nonumber\\
&&\;\;\;\;\;\;\;\;\;\;\;\;\;\;\;\;\;{\rm{                       0}} \le {P_{m}} \le P_{\rm{NGSO2}}^{\max }\nonumber\\
&&\;\;\;\;\;\;\;\;\;\;\;\;\;\;\;\;\;{\rm{                       0}} \le {P_{i}} \le P_{\rm{BS}}^{\max }\nonumber\\
&&\;\;\;\;\;\;\;\;\;\;\;\;\;\;\;\;\;{\rm{                       SINR}}_{{{\rm{U}}_{\rm{N2}}^j}} \ge \phi _{th}\nonumber\\
&&\;\;\;\;\;\;\;\;\;\;\;\;\;\;\;\;\;{\rm{                       SINR}}_{{{\rm{U}}_{\rm{BS}}^t}} \ge \phi _{th}\nonumber\\
&&\;\;\;\;\;\;\;\;\;\;\;\;\;\;\;\;\;{\rm{                       }}{\alpha _n},{\alpha _m} \in \left\{ {0,1} \right\}, \label{equ:10}
\end{eqnarray}
where $P_{\rm{NGSO1}}^{\max }$, $P_{\rm{NGSO2}}^{\max }$, and $P_{\rm{BS}}^{\max }$ are the maximum available transmit power of NGSO 1, of NGSO 2, and of the BSs, respectively, while $\phi _{\rm{th}}$ is the minimum SINR required.

Note that optimizing the the $k$-th user's performance is expected to affect the interference imposed on its neighbors in NGSO 1. However, as mentioned, the effect of the intra-constellation CFI may be negligible in the presence of the multi-color multibeam frequency-reuse schemes (e.g., FR7) as shown in Fig. 3. More explicitly, optimizing the SINR of a user $k$ does not substantially increase the interference imposed on users at the same frequency of a user $k$ in NGSO 1. Similarly, the SINR of a user $k$ may not be obviously affected when optimizing the SINR of other users accessing the same frequency of this user $k$. Thus, although there may exist a number of users in NGSO 1, the SINR of other users can be optimized similarly to the $k$-th user, $1 \le k \le K$.

This optimization problem is related with the coverage of a pair of NGSO constellations. In the overlapping coverage, given the tele-traffic states, the CSI, and the transmission requirements, the achievable SINR depends both on the beam-scheduling and power-scheduling. It is challenging to directly solve such a complex problem, which hinges on both the coverage analysis and beam-power scheduling.

%\IEEEpubidadjcol

\section{Jointly Utilizing Multi-Domain Resources for Interference Management}
In this section, we present our multi-domain resource aided interference management scheme, including both beam shut-off and beam switching based scheduling, as well as our LSTM-ARMA-DQN based power allocation.

\subsection{Shut-Off and Switching-Based Beam Scheduling}
In this subsection, we conceive a shut-off and switching-based beam scheduling method for managing the CFI. Specifically, using the Poisson traffic-generation model of \cite{BIBPoisson} for NGSO 1 and NGSO 2, the tele-traffic states of the satellites in ${\Phi _{{{\rm{N2}}_m} \to {{\rm{U}}_{\rm{N1}}^k}}}$, and ${\Phi _{{{\rm{N2}}_m} \to {{\rm{U}}_{\rm{N2}}^j}}}$, can be readily estimated. Given on the estimated tele-traffic states, the idle beams of ${{\rm{N1}}_n}$ and of ${{\rm{N2}}_m}$ are shut off for reducing the amount of interference imposed on NGSO 1's users. As for the active beams of the satellites in ${\Phi _{{{\rm{N2}}_m} \to {{\rm{U}}_{\rm{N1}}^k}}}$, we perform beam switching. More specifically, if there is a satellite ${{\rm{N2}}_{m^{'}}}, m^{'}\ne m$ in ${\Phi _{{{\rm{N2}}_m} \to {{\rm{U}}_{\rm{N2}}^j}}}$ having an idle co-frequency beam, which is not covering ${{\rm{U}}_{\rm{N1}}^k}$ (${{\rm{N2}}_{m^{'}}} \notin {\Phi _{{{\rm{N2}}_m} \to {{\rm{U}}_{\rm{N1}}^k}}}$), the ${{\rm{N2}}_m}$ forwards its traffic to this ${{\rm{N2}}_{m^{'}}}$ and informs ${{\rm{U}}_{\rm{N2}}^j}$. Then, ${{\rm{N2}}_{m^{'}}}$ may point its beam to ${{\rm{U}}_{\rm{N2}}^j}$, and sends the data received from ${{\rm{N2}}_m}$ to ${{\rm{U}}_{\rm{N2}}^j}$. As expected, the beam switching action of NGSO 1 is similar to that of NGSO 2. Additionally, the specific procedure of the shut-off and switching based beam scheduling designed is summarized in Algorithm 1.

\begin{table}[htbp]
\centering
\small
\begin{tabular}{p{0.45\textwidth}}
\bottomrule
\textbf{Algorithm 1: Beam Shut-Off and Beam Switching}\\
\hline
\textbf{Input:} ${\Phi _{{{\rm{N2}}_m} \to {{\rm{U}}_{\rm{N1}}^k}}}$, ${\Phi _{{{\rm{N2}}_m} \to {{\rm{U}}_{\rm{N2}}^j}}}$, ${\Phi _{{{\rm{N1}}_n} \to {{\rm{U}}_{\rm{N2}}^j}}}$, ${\Phi _{{{\rm{N1}}_n} \to {{\rm{U}}_{\rm{BS}}^t}}}$ and ${\Phi _{{{\rm{N2}}_m} \to {{\rm{U}}_{\rm{BS}}^t}}}$, $p({{\rm{N2}}_m})$, $p({{\rm{N1}}_n})$, $p({{\rm{U}}_{\rm{N1}}^k})$, $p({{\rm{U}}_{\rm{N2}}^j})$, $p({{\rm{BS}}_i})$, $p({{\rm{U}}_{\rm{BS}}^t})$ and the traffic states of the ${{\rm{N1}}_n}$ and ${{\rm{N2}}_m}$.\\
1. Turn off the co-frequency beams of the satellites in ${\Phi _{{{\rm{N2}}_m} \to {{\rm{U}}_{\rm{N1}}^k}}}$ and ${\Phi _{{{\rm{N1}}_n} \to {{\rm{U}}_{\rm{N2}}^j}}}$, if these beams have no active traffic in a given time slot.\\
2. Remove the satellites from ${\Phi _{{{\rm{N2}}_m} \to {{\rm{U}}_{\rm{N1}}^k}}}$ and ${\Phi _{{{\rm{N1}}_n} \to {{\rm{U}}_{\rm{N2}}^j}}}$, if their beams are turned off in Step 1.\\
3. If idle beams of the satellites in ${\Phi _{{{\rm{N2}}_m} \to {{\rm{U}}_{\rm{N2}}^j}}}$ are available, and there is one idle beam without covering ${{\rm{U}}_{\rm{N1}}^k}$, then\\
4. \;\;\;\;Switch the interfering beams of the satellites in ${\Phi _{{{\rm{N2}}_m} \to {{\rm{U}}_{\rm{N1}}^k}}}$ having traffic to idle beams of the satellites in ${\Phi _{{{\rm{N2}}_m} \to {{\rm{U}}_{\rm{N2}}^j}}}$.\\
5. \;\;\;\;Remove the satellites from ${\Phi _{{{\rm{N2}}_m} \to {{\rm{U}}_{\rm{N1}}^k}}}$, if their beams are switched in Step 4.\\
6. End\\
7. If idle beams of the satellites in ${\Phi _{{{\rm{N1}}_n} \to {{\rm{U}}_{\rm{N1}}^k}}}$ are available, and there is one idle beam without covering ${{\rm{U}}_{\rm{N2}}^j}$, then\\
8. \;\;\;\;Switch the interfering beams of the satellites in ${\Phi _{{{\rm{N1}}_{n^{'}}} \to {{\rm{U}}_{\rm{N2}}^j}}}$, $n^{'} \ne n$, having active traffic to idle beams of satellites, similarly to Step 4.\\
9. End\\
10. Obtain the updated ${\Phi _{{{\rm{N2}}_m} \to {{\rm{U}}_{\rm{N1}}^k}}}$ and ${\Phi _{{{\rm{N1}}_n} \to {{\rm{U}}_{\rm{N2}}^j}}}$.\\
11. Obtain the updated ${\Phi _{{{\rm{N1}}_n} \to {{\rm{U}}_{\rm{BS}}^t}}}$ and ${\Phi _{{{\rm{N2}}_m} \to {{\rm{U}}_{\rm{BS}}^t}}}$.\\
12. Repeat Step 1$\sim$7 for different ${{{\rm{N1}}_n}}$, and its users.\\
\textbf{Output:} The number of interfering satellites $N_{\rm{S1}}^{'}$, $N_{{\rm{S2}}}^{'}$, $N_{{\rm{S3}}}^{'}$ and $N_{{\rm{S4}}}^{'}$.\\
\toprule
\end{tabular}
\end{table}

Furthermore, the optimization problem (10), can be simplified as
\begin{eqnarray}
&&\mathop {\max }\limits_{{P_{n}},{P_{m}},{P_{i}}} {\rm{SINR}}_{{{\rm{U}}_{\rm{N1}}^k}}^{'}\nonumber \\
&&{\rm{         s.t.}}\;\;\;\;\;\;\;\;\;\;\;\;{{         0}} \le {P_{n}} \le P_{\rm{NGSO1}}^{\max }\nonumber\\
&&\;\;\;\;\;\;\;\;\;\;\;\;\;\;\;\;\;{{                       0}} \le {P_{m}} \le P_{\rm{NGSO2}}^{\max }\nonumber\\
&&\;\;\;\;\;\;\;\;\;\;\;\;\;\;\;\;\;{{                       0}} \le {P_{i}} \le P_{\rm{BS}}^{\max }\nonumber\\
&&\;\;\;\;\;\;\;\;\;\;\;\;\;\;\;\;\;{\rm{                       SINR}}_{{{\rm{U}}_{\rm{N2}}^j}}^{'} \ge \phi _{\rm{th}}\nonumber\\
&&\;\;\;\;\;\;\;\;\;\;\;\;\;\;\;\;\;{\rm{                       SINR}}_{{{\rm{U}}_{\rm{BS}}^t}}^{'} \ge \phi _{\rm{th}}, \label{equ:11}
\end{eqnarray}
where ${\rm{SINR}}_{{{\rm{U}}_{\rm{N1}}^k}}^{'}$, ${\rm{ SINR}}_{{{\rm{U}}_{\rm{N2}}^j}}^{'}$ and ${\rm{ SINR}}_{{{\rm{U}}_{\rm{BS}}^t}}^{'}$ are $\frac{{{P_{n}}G_{n{{\rm{U}}_{\rm{N1}}^k}}^tG_{n{{\rm{U}}_{\rm{N1}}^k}}^rL_{nk}^S}{{| {{h_{nk}}} |}^2}}{{\sum\limits_{m = 1}^{{N_{{\rm{S1}}}^{'}}}{{P_{m}}G_{m{{\rm{U}}_{\rm{N1}}^k}}^tG_{m{{\rm{U}}_{\rm{N1}}^k}}^rL_{mk}^S}{{| {{h_{mk}}} |}^2} + \sum\limits_{i \in \Phi _{\rm{BS}}}\!\!\!{{P_{i}}L_{ik}^S}{{| {{h_{ik}}} |}^2} + k_BB_nT_n}}$, $ \frac{{{P_{m}}G_{m{{\rm{U}}_{\rm{N2}}^j}}^tG_{m{{\rm{U}}_{\rm{N2}}^j}}^rL_{mj}^S}{{| {{h_{mj}}} |}^2}}{{\sum\limits_{n = 1}^{{N_{{\rm{S2}}}^{'}}}{{P_{n}}G_{n{{\rm{U}}_{\rm{N2}}^j}}^tG_{n{{\rm{U}}_{\rm{N2}}^j}}^rL_{nj}^S}{{| {{h_{nj}}} |}^2} +\!\!\! \sum\limits_{i \in \Phi _{\rm{BS}}}\!\!\!\!\!\!{{P_{i}}L_{ij}^S}{{| {{h_{ij}}} |}^2} + k_BB_mT_n}}$, and $\frac{{{P_{i}}L_{it}^S}{{| {{h_{it}}} |}^2}}{{\sum\limits_{n = 1}^{{N_{{\rm{S3}}}^{'}}}{{P_{n}}L_{nt}^S}{{| {{h_{nt}}} |}^2} + \sum\limits_{m = 1}^{{N_{{\rm{S4}}}^{'}}}{{P_{m}}L_{mt}^S}{{| {{h_{mt}}} |}^2}+\!\!\!\!\!\!\!\!\!\sum\limits_{z \in \Phi _{\rm{BS}}, z \ne i}\!\!\!\!\!\!\!{{P_{z}}L_{zt}^S}{{| {{h_{zt}}} |}^2} + \sigma _n^2}}$, respectively.

\subsection{LSTM-ARMA-DQN Based Power Allocation}
As shown in Fig.~\ref{fig:fig6}, a LSTM-ARMA-DQN neural network, combining a LSTM-ARMA and a DQN, is designed for assisting power allocation. Such a LSTM-ARMA model is constructed for predicting the instantaneous CSIs of the links spanning from the satellites to the terrestrial users, using the historical CSIs estimated with the aid of \cite{BIB25}. Explicitly, the LSTM is used, since it does well in time series prediction \cite{BIBCSIP}. To further improve the prediction performance of the LSTM, the ARMA is used for optimizing the prediction error \cite{BIBour,BIB28}, and for upgrading the loss of the LSTM in the model-training stage, since the ARMA is proficient at evaluating the prediction error \cite{BIBOURARMA} relying on the parameters of the autoregressive model and the moving average model \cite{BIB26}. Using the upgraded loss and the back propagation algorithm \cite{BIBBP}, the LSTM can be further optimized. Moreover, the predicted CSIs may be entered into the DQN constructed \cite{BIB27}. Following our previous work \cite{BIBour,BIB28}, the model constructed is optimized with the aid of grid search and K-fold cross-validation. Specifically, the CSI dataset is first divided into K subsets. Then, given the $k$-th K-fold Cross-Validation, the $k$-th subset and the remaining subsets are used for validating and training the constructed model, respectively. Additionally, the parameters of the LSTM-ARMA-DQN are optimized with the aid of the adaptive moment estimation optimizer of \cite{BIB28}. The optimized LSTM is composed by two LSTM layers. Their size is 128 and 64. Meanwhile, the optimized DQN is composed by six fully connected (FC) layers, wherein the Rectified Linear Unit function is used as the activation function for each layer of the first three FC layers, and the Hyperbolic Tangent function is utilized as the activation function for each layer of the last three FC layers. The size of six FC layers are 256, 256, 512, 512, 1024 and 1024. Then, the LSTM-ARMA-DQN model is used for optimizing the ${P_{n}}$, ${P_{m}}$ and ${P_{i}}$. The specific optimization procedure is summarized in Algorithm 2.

\begin{figure}[htbp]
\centering
\includegraphics[width=0.9\linewidth]{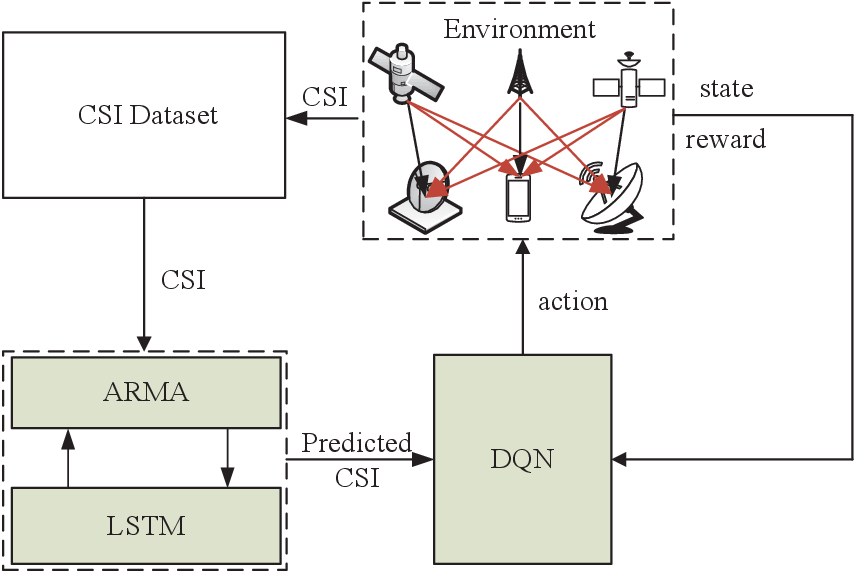}
\caption{LSTM-ARMA-DQN aided power allocation.\label{fig:fig6}}
\end{figure}

\begin{table}[htbp]
\centering
\small
\begin{tabular}{p{0.45\textwidth}}
\bottomrule
\textbf{Algorithm 2: LSTM-ARMA-DQN Based Power Allocation}\\
\hline
\textbf{Input:} ${\Phi _{{\rm{N2}}_m} \to {{\rm{U}}_{\rm{N1}}^k}}$, ${\Phi _{{\rm{N2}}_m} \to {{\rm{U}}_{\rm{N2}}^j}}$, ${\Phi _{{\rm{N1}}_n} \to {{\rm{U}}_{\rm{N1}}^k}}$, ${\Phi _{{\rm{N1}}_n} \to {{\rm{U}}_{\rm{N2}}^j}}$, ${\Phi _{{\rm{N2}}_m} \to {{\rm{U}}_{\rm{BS}}^t}}$, ${\Phi _{{\rm{N1}}_n} \to {{\rm{U}}_{\rm{BS}}^t}}$, $p({{\rm{N2}}_m})$, $p({{\rm{N1}}_n})$, $p({{\rm{BS}}_i})$, $p({{\rm{U}}_{\rm{N1}}^k})$, $p({{\rm{U}}_{\rm{N2}}^j})$, $p({{\rm{U}}_{\rm{BS}}^t})$, $P_{\rm{NGSO1}}^{\max }$, $P_{\rm{NGSO2}}^{\max }$, $P_{\rm{BS}}^{\max }$, and $\phi _{th}$.\\
1. Initialize the parameters of the LSTM-ARMA-DQN model constructed, the transmit power vector $\mathbf P_n$, the SINR vector $\mathbf S_{\rm{SINR}}$, the SINR threshold ${\rm{SINR}}_{{{{\rm{U}}_{\rm{N1}}}}}^{max}$, the greedy probability threshold $\varepsilon$-greedy.\\
2. Construct the LSTM-ARMA-DQN model, and train the LSTM-ARMA model constructed.\\
3. Calculate the antenna gain $G_{ab}^d$, using the input satellite sets and the locations.\\
4. Predict the CSIs using the well-trained LSTM-ARMA model at a given transmission slot.\\
5. Calculate the ${\rm{SINR}}_{{{\rm{U}}_{\rm{BS}}^t}}^{'}$, and compare it with $\phi _{th}$ to adjust $P_i$, and calculate the required $P_m$ according to the transmission specifications.\\
6. Calculate the ${\rm{SINR}}_{{{\rm{U}}_{\rm{N1}}^k}}^{'}$ using the initialized $P_n$, the obtained $P_i$ and $P_m$, and assign the calculated ${\rm{SINR}}_{{{\rm{U}}_{\rm{N1}}^k}}^{'}$ to $\mathbf S_{\rm{SINR}}$.\\
7. Set a positive reward if the calculated ${\rm{SINR}}_{{{\rm{U}}_{\rm{N1}}^k}}^{'} \ge {\rm{SINR}}_{{{{\rm{U}}_{\rm{N1}}}}}^{max}$ and $P_m \le P_{\rm{NGSO2}}^{\max }$, otherwise give a negative reward.\\
8. Select the max value of the DQN output, assign the index of the selected value to INDEX, if RAND(0,1)$>\varepsilon$-greedy. Otherwise, set INDEX to a random integer value.\\
9. Update $P_n$ by selecting a value from $\mathbf P_n$ using INDEX.\\
10. Repeat Step 5 $\sim$ 9, until the maximum number of iterations is reached.\\
11. Obtain the optimized $P_n$, $P_m$ and $P_i$ using $\mathbf S_{\rm{SINR}}$, and repeat Step 3 $\sim$ 10 for different ${{{\rm{N1}}_n}}$.\\
\textbf{Output:} The optimized $P_n^{o}$, $P_m^{o}$, and $P_i^{o}$.\\
\toprule
\end{tabular}
\end{table}

We should point out that our deep-learning-based power allocation algorithm can be used to perform power allocation for the SGIN considered, without requiring channel estimation. In our algorithm, the deep-learning model conceived can predict the CSI, and allocates the transmit power to the SGIN considered. More specifically, due to the propagation delay and the potential co-frequency interference among users, the estimated CSI may become outdated and imperfect. As a remedy, this deep-learning-based power allocation algorithm may be adopted. With the improvement of the computing capability and the multi-access edge computing technology on board satellites\cite{BIBMEC1,BIBMEC2}, this algorithm may be performed on satellites, to avoid the propagation delay between the satellites and the gateways, if the algorithm is executed in the gateways.

\section{Outage Probability Analysis for The Proposed JMDR-IM Scheme}
In this section, we derive the OP of the proposed JMDR-IM scheme. Moreover, we also provide the asymptotic OP analysis of the proposed JMDR-IM scheme.

Using \cite{BIB29}, the definition of the OP is given by
\begin{equation}
{P_{\rm{OP}}} = \Pr \left( {\rm{SINR}} < {\phi _{\rm{th}}} \right). \label{equ:12}
\end{equation}

\subsection{NGSO 1 Outage Probability Analysis}

Using (4) and (12), the OP of the ${\rm{N1}}_n$-${{\rm{U}}_{\rm{N1}}^k}$ link in the proposed JMDR-IM scheme is written as
\begin{eqnarray}
&&\!\!\!\!\!\!\!\!\!\!\!\!\!{P_{\rm{OP}}^{\rm{NGSO1}}} \nonumber\\ &&\!\!\!\!\!\!\!\!\!\!\!\!\!=\!\Pr\!\left(\!\frac{{{P_{n}^o}G_{n{{\rm{U}}_{\rm{N1}}^k}}^tG_{n{{\rm{U}}_{\rm{N1}}^k}}^rL_{nk}^S}{{\left| {{h_{nk}}} \right|}^2}}{{\sum\limits_{m = 1}^{{N_{{\rm{S1}}}^{'}}}\!\!{{P_{m}^o}\!G_{m\!{{\rm{U}}_{\rm{N1}}^k}}^t\!\!\!G_{m\!{{\rm{U}}_{\rm{N1}}^k}}^r\!\!L_{mk}^S}{{\left| {{h_{mk}}} \right|}^2} \!\!+\!\!\!\! \sum\limits_{i \in \Phi _{\rm{BS}}}\!\!\!\!{{P_{i}^o}L_{ik}^S}{{\left| {{h_{ik}}} \right|}^2} \!+\! \sigma _{\rm{N1}}^2}}\right.\nonumber\\
&&<{\phi _{\rm{th}}}\Bigg), \label{equ:13}
\end{eqnarray}
where we have $\sigma _{\rm{N1}}^2 = k_BB_nT_n$, $L_{ik}^S=L_{i{{\rm{U}}_{\rm{N1}}^k}}^{-\alpha _t}$, ${\alpha _t}$ is the path loss exponent, and $L_{i{{\rm{U}}_{\rm{N1}}^k}}$ is the distance between the ${\rm{BS}}_i$ and the ${\rm{U}}_{\rm{N1}}^k$.

Using (A.7), ${P_{\rm{OP}}^{\rm{NGSO1}}}$ can be finally obtained.

To further explore the OP of the ${\rm{N1}}_n$-${{\rm{U}}_{\rm{N1}}^k}$ link in the proposed JMDR-IM scheme, we analyze the asymptotic outage probability (AOP). The PDF of ${P_n}{{| {{h_{nk}}} |}^2}$ can be formulated as
\begin{equation}
{f_{{P_n}{{\left|  {{h_{nk}}}  \right|}^2}}}\left( x \right) \!=\! \frac{\alpha _{{\rm{N1}}_n}}{P_n}{e^{ - {\beta _{{\rm{N1}}_n}}\frac{x}{P_n}}}{}_1{F_1}\!\left(\! {{m_{{\rm{N1}}_n}};1;{\delta _{{\rm{N1}}_n}}\frac{x}{P_n}} \!\right).\label{equ:14}
\end{equation}

For a large ${P_n}$, we have ${}_1{F_1}( {{m_{{\rm{N1}}_n}};1;{\delta _{{\rm{N1}}_n}}\frac{x}{P_n}} )\to 1$ \cite{BIB29}. Thus, ${f_{{P_n}{{|  {{h_{nk}}}  |}^2}}}( x )$ can be rewritten as
\begin{equation}
{f_{{P_n}{{\left|  {{h_{nk}}}  \right|}^2}}}\left( x \right) = \frac{\alpha _{{\rm{N1}}_n}}{P_n}{e^{ - {\beta _{{\rm{N1}}_n}}\frac{x}{P_n}}}.\label{equ:15}
\end{equation}

Using (15), similarly to (A.7), the AOP of the ${\rm{N1}}_n$-${{\rm{U}}_{\rm{N1}}^k}$ link in the proposed JMDR-IM scheme can be expressed as
\begin{eqnarray}
&&\!\!\!\!\!\!\!\!\!\!\!\!\!\!\!\!\!\!\!\!\!{P_{\rm{AOP}}^{\rm{NGSO1}}}\!=\! \frac{{{\alpha _{{\text{N}}_{{\text{1}}{{n}}}}}}}{{{\beta _{{\rm{N}}_{{1_n}}}}}}\!\prod\limits_{m \!=\! 1}^{{{N}}_{{\rm{S1}}}^{'}}\! {\sum\limits_{{n_{{\rm{N}}_{{2_m}}}} \!=\! 0}^{{m_{{{\rm{N}}_{{2_m}}}}}-1}\! {Z_{{{\rm{N2}}_m}}} }- \frac{{{\alpha _{{\text{N}}_{{\text{1}}{\text{n}}}}}}}{{{\beta _{{\rm{N}}_{{1_n}}}}}}{e^{ - \frac{{{\beta _{{\rm{N}}_{{1_n}}}}}}{{{P_n}}}{S^{'}}\sigma _{{\rm{N}}1}^2}}\nonumber\\
&&\;\;\;\; \prod\limits_{m = 1}^{{{N}}_{{\rm{S1}}}^{'}} {\sum\limits_{{n_{{\rm{N}}_{{2_m}}}} = 0}^{{m_{{{\rm{N}}_{2_m}}}} - 1} }\frac{{Z_{{{\rm{N2}}_m}}}\left({\beta _{{\rm{N}}_{{2_m}}}} \!-\! {\delta _{{{\rm{N}}_{{2_m}}}}}\right)}{\left( {{\beta _{{\rm{N}}_{{1_n}}}} \!-\! {\delta _{{{\rm{N}}_{{1_n}}}}} \!+\! \frac{{{\beta _{{\rm{N}}_{{1_n}}}}}}{{{P_n}}}\zeta _m^{'}} \right)^{ {n_{{\rm{N}}_{{2_m}}}} + 1}}\nonumber\\
&&\;\;\;\;\;\;\;\;\;\;\;\;\;\;\;\;\;\;\;\;\;\prod\limits_{i = 1}^{{{\rm{N}}_{{\rm{BS}}}}} {\frac{1}{{\sigma _{{\rm{BS}}}^2\left( {\frac{{{\beta _{{\rm{N}}_{{1_n}}}}}}{{{P_n}}}{\rm{I}}_{{\rm{BS}}}^{'}  + \frac{1}{{\sigma _{{\rm{BS}}}^2}}} \right)}}} ,\label{equ:16}
\end{eqnarray}
where $\zeta _m^{'} = \frac{{{\phi _{\rm{th}}}G_{m{\text{U}}_{{\rm{N1}}}^k}^tG_{m{\text{U}}_{{\rm{N1}}}^k}^rL_{mk}^S{P_m}}}{{G_{n{\text{U}}_{{\rm{N1}}}^k}^tG_{n{\text{U}}_{{\rm{N1}}}^k}^rL_{nk}^S}}$, ${S^{'}} = \frac{{{\phi _{th}}}}{{G_{n{\text{U}}_{{\rm{N1}}}^k}^tG_{n{\text{U}}_{{\rm{N1}}}^k}^rL_{nk}^S}}$ and ${\rm{I}}_{{\rm{BS}}}^{'} = \frac{{{\phi _{{\rm{th}}}}{P_i}L_{ik}^S}}{{G_{n{\text{U}}_{{\rm{N}}1}^k}^tG_{n{\text{U}}_{{\rm{N}}1}^k}^rL_{nk}^S}}$.

We should point out that the OP of NGSO 1 can be dramatically reduced. This beneficial result is achieved because the SINR of NGSO 1 can be increased by the JMDR-IM scheme, and an increased SINR results in an OP reduction.

\subsection{NGSO 2 Outage Probability Analysis}
Moreover, using (5) and (12), the OP of the ${\rm{N2}}_m$-${{\rm{U}}_{\rm{N2}}^j}$ link in the proposed JMDR-IM scheme is formulated as
\begin{eqnarray}
&&\!\!\!\!\!\!\!\!\!\!\!\!\!{P_{\rm{OP}}^{\rm{NGSO2}}} \nonumber\\ &&\!\!\!\!\!\!\!\!\!\!\!\!\!=\!\Pr\!\left(\!\frac{{{P_{m}^o}G_{m{{\rm{U}}_{\rm{N2}}^j}}^tG_{m{{\rm{U}}_{\rm{N2}}^j}}^rL_{mj}^S}{{\left| {{h_{mj}}} \right|}^2}}{{\sum\limits_{n = 1}^{{N_{{\rm{S2}}}^{'}}}\!{{P_{n}^o}G_{\!n{{\rm{U}}_{\rm{N2}}^j}}^t\!\!G_{\!\!n{{\rm{U}}_{\rm{N2}}^j}}^r\!\!\!L_{nj}^S}{{\left| {{h_{nj}}} \right|}^2}\!\! +\!\!\! \sum\limits_{i \in \Phi _{\rm{BS}}}\!\!\!\!{{P_{i}^o}L_{ij}^S}{{\left| {{h_{ij}}} \right|}^2} \!+\! \sigma _{\rm{N2}}^2}}\right.\nonumber\\
&& <{\phi _{th}}\Bigg),\label{equ:17}
\end{eqnarray}
where $\sigma _{\rm{N2}}^2 = k_BB_mT_n$, $L_{ij}^S=L_{i{{\rm{U}}_{\rm{N2}}^j}}^{-\alpha _t}$ and $L_{i{{\rm{U}}_{\rm{N2}}^j}}$ is the distance between ${\rm{BS}}_i$ and ${\rm{U}}_{\rm{N2}}^j$.

Similarly to (A.7), ${P_{\rm{OP}}^{\rm{NGSO2}}}$ can be formulated as
\begin{eqnarray}
&&\!\!\!\!\!\!\!\!\!\!\!\!\!\!\!\!\!\!\!\!{P_{\rm{OP}}^{\rm{NGSO2}}}\nonumber\\
&&\!\!\!\!\!\!\!\!\!\!\!\!\!\!\!\!\!\!\!\! =\!\!\! \sum\limits_{{n_{{\rm{N}}{2_m} = 0}}}^{{m_{{\rm{N}}{2_m} \!-\! 1}}} \!{\left(\!\! {{Z_{{\rm{N}}{2_m}}} \!\!\prod\limits_{n = 1}^{{{N}}_{{\rm{S2}}}^{'}}\! \sum\limits_{{n_{{\rm{N}}{1_n}}} = 0}^{{m_{{\rm{N}}{1_n}}}\! - \!1}\!\!\! {Z_{{{\rm{N1}}_n}}}} \right.}\!\!\!-\!\! {Z_{{\rm{N}}_{{2_m}}}}\!\!{e^{ -\! \left(\! {{\beta _{{{\rm{N2}}_m}}} \!\!-\! {\delta _{{{{\rm{N2}}_m}}}}} \!\right)S\sigma _{{\rm{N2}}}^2}}  \nonumber \\
&&\sum\limits_{{k_o} \!=\! 0}^{{n_{{\rm{N}}{1_n}}}}\! {\sum\limits_{{j_o} \!=\! 0}^{{k_o}}\! {\sum\limits_{{z_o\!=\!0}}^{{j_o}}\! {\frac{{{{\left(\! {{\beta _{{{\rm{N}}_{{2_m}}}}}\!\!\!\! -\! {\delta _{{{{\rm{N}}_{{2_m}}}}}}} \!\right)}^{{k_o}}}}}{{{k_o}!}}} } }\left(\! {\begin{matrix}
   {{k_o}}  \\
   {{j_o}}  \\
 \end{matrix} } \!\right)\!\left(\! {\begin{matrix}
   {{j_o}}  \\
   {{z_o}}  \\
 \end{matrix} } \!\right)\!{{\left(\! {{S_1}\sigma _{{\rm{N}}2}^2} \!\right)}^{{j_o} \!-\! {z_o} }} \nonumber\\
&&\sum\limits_{x_n\ge0,{x_1},{x_2},\cdots,{x_{{{N}}_{{\rm{S2}}}^{'}}}}^{{k_o\!-\!j_o}}\!{\left(\! {\begin{matrix}
   {{k_o\!-\!j_o}}  \\
   {{x_1},\!{x_2},\cdots,{x_{{{N}}_{{\rm{S2}}}^{'}}}}  \\
 \end{matrix} } \!\right)\!\prod\limits_{n = 1}^{{{N}}_{{\rm{S2}}}^{'}}\! {{{\left( \!{{\zeta _n}} \!\right)}^{{x_n}}}} }\nonumber\\
&& \prod\limits_{n = 1}^{{{N}}_{{\rm{S2}}}^{'}} {\left(\! {\sum\limits_{{n_{{\rm{N}}{1_n}}}=0}^{{m_{{\rm{N}}{1_n}}} \!-\! 1}\!\! {Z_{{\rm{N}}_{{1_n}}}}\left(\! {{\beta _{{\rm{N}}{1_n}}} \!\!-\!\! {\delta _{{{\rm{N}}{1_n}}}}} \!\right) \left(\! {{x_n} \!\!+\!\! {n_{{\rm{N}}{1_n}}}} \!\right)!} \right.} \nonumber\\
&&{\left( {\left( {{\beta _{{{\rm{N}}_{{2_m}}}}}- {\delta _{{{{\rm{N}}_{{2_m}}}}}}} \right){\zeta _n}{{ - }}\left(\! {{\beta _{{\rm{N}}{1_n}}} \!-\! {\delta _{{{\rm{N}}{1_n}}}}} \right)} \right)^{ - {x_n} - {n_{{\rm{N}}{1_n}}}-1}} \nonumber\\
&&\sum\limits_{y_i\!\ge\!0,{y_1},{y_2},\!\cdots\!,{y_{{{\rm{N}}_{{\rm{BS}}}}}}}^{{z_o}} {{z_y}\prod\limits_{i = 1}^{{{\rm{N}}_{{\rm{BS}}}}}\! {{{\left( {{I_{{\rm{BS}}{_i}}}} \right)}^{{y_i}}}} } \!\prod\limits_{i = 1}^{{{\rm{N}}_{{\rm{BS}}}}}  {\frac{1}{{\sigma _{{\rm{BS}}}^2}}\left( {{y_i}} \right)!}  \nonumber\\
&&\left. {\left. {{{\left( {\left( {{\beta _{{{\rm{N}}_{{2_m}}}}} - {\delta _{{{{\rm{N}}_{{2_m}}}}}}} \right){I_{{\rm{BS1}}{_i}}} + \frac{1}{{\sigma _{{\rm{BS}}}^2}}} \right)}^{ - {y_i} - 1}}} \right)} \right),\label{equ:18}
\end{eqnarray}

where we have ${Z_{{{\rm{N2}}_m}}} = \frac{{{\alpha _{{{\rm{N2}}_m}}}( {{m_{{{{\rm{N2}}_m}}}} - 1} )!{{( {{\delta _{{{{\rm{N2}}_m}}}}} )}^{{n_{{{{\rm{N2}}_m}}}}}}}}{{( {{m_{{{{\rm{N2}}_m}}}} - 1 - {n_{{{{\rm{N2}}_m}}}}} )!{{( 1 )}_{{n_{{{{\rm{N2}}_m}}}}}}{{( {{\beta _{{{\rm{N2}}_m}}} - {\delta _{{{{\rm{N2}}_m}}}}} )}^{{n_{{{\rm{N2}}_m}}} + 1}}}}$, ${\zeta _n} = \frac{{{\phi _{\rm{th}}}P_n^oG_{n{\text{U}}_{{\rm{N2}}}^j}^tG_{n{\text{U}}_{{\rm{N2}}}^j}^rL_{nj}^s}}{{P_m^oG_{m{\text{U}}_{{\rm{N2}}}^j}^tG_{m{\text{U}}_{{\rm{N2}}}^j}^rL_{mj}^s}}$, ${I_{{\rm{BS1}}_i}} = \frac{{{\phi _{\rm{th}}}P_i^oL_{ij}^s}}{{P_m^oG_{m{\text{U}}_{{\rm{N2}}}^j}^tG_{m{\text{U}}_{{\rm{N2}}}^j}^rL_{mj}^s}}$, and $S_1 = \frac{{{\phi _{\rm{th}}}}}{{P_m^oG_{m{\text{U}}_{{\rm{N2}}}^j}^tG_{m{\text{U}}_{{\rm{N2}}}^j}^rL_{mj}^s}}$.

To evaluate the OP of a large $P_m$, similarly to (16), the AOP of the ${\rm{N2}}_m$-${{\rm{U}}_{\rm{N2}}^j}$ link in the proposed JMDR-IM scheme can be expressed as
\begin{eqnarray}
&&\!\!\!\!\!\!\!\!\!\!\!\!\!\!\!\!\!\!\!\!\!{P_{\rm{AOP}}^{\rm{NGSO2}}}\!\!=\!\! \frac{{{\alpha _{{\text{N}}_{{\text{2}}{{m}}}}}}}{{{\beta _{{\rm{N}}_{{2_m}}}}}}\!\!\prod\limits_{n \!=\! 1}^{{\rm{N}}_{{\rm{S2}}}^{'}}\!\! {\sum\limits_{{n_{{\rm{N}}_{{1_n}}}} \!=\! 0}^{{m_{{{\rm{N}}_{{1_n}}}}}-1}\!\!\! {Z_{{{\rm{N1}}_n}}} }-\frac{{{\alpha _{{\text{N}}_{{\text{2}}{{m}}}}}}}{{{\beta _{{\rm{N}}_{{2_m}}}}}}{e^{ - \frac{{{\beta _{{\rm{N}}_{{2_m}}}}}}{{{P_m}}}{{S1}^{'}}\sigma _{{\rm{N}}2}^2}}\nonumber\\
&&\;\;\;\;  \prod\limits_{n = 1}^{{\rm{N}}_{{\rm{S2}}}^{'}} {\sum\limits_{{n_{{\rm{N}}_{{1_n}}}} = 0}^{{m_{{{\rm{N}}_{{1_n}}}}} - 1} {Z_{{{\rm{N1}}_n}}}\left({{\beta _{{\rm{N}}_{{1_n}}}} \!-\! {\delta _{{{\rm{N}}_{{1_n}}}}}}\right) } \nonumber\\
&&\;\;\;\;\;\;{\left( {{\beta _{{\rm{N}}_{{2_m}}}} \!-\! {\delta _{{{\rm{N}}_{{2_m}}}}} \!+\! \frac{{{\beta _{{\rm{N}}_{{2_m}}}}}}{{{P_m}}}\zeta _n^{'}} \right)^{ - {n_{{\rm{N}}_{{1_n}}}} - 1}}\nonumber\\
&&\;\;\;\;\;\;\prod\limits_{i = 1}^{{{\rm{N}}_{{\rm{BS}}}}} {\frac{1}{{\sigma _{{\rm{BS}}}^2\left( {\frac{{{\beta _{{\rm{N}}_{{2_m}}}}}}{{{P_m}}}{\rm{I}}_{{\rm{BS1}}}^{'} + \frac{1}{{\sigma _{{\rm{BS}}}^2}}} \right)}}} ,\label{equ:19}
\end{eqnarray}

where $\zeta _n^{'} = \frac{{{\phi _{{\rm{th}}}}G_{n{\text{U}}_{{\rm{N}}2}^j}^tG_{n{\text{U}}_{{\rm{N}}2}^j}^rL_{nj}^S{P_n}}}{{G_{m{\text{U}}_{{\rm{N}}2}^j}^tG_{m{\text{U}}_{{\rm{N}}2}^j}^rL_{mj}^S}}$, ${{S1}^{'}} = \frac{{{\phi _{{\rm{th}}}}}}{{G_{m{\text{U}}_{{\rm{N}}2}^j}^tG_{m{\text{U}}_{{\rm{N}}2}^j}^rL_{mj}^S}}$, and ${\rm{I}}_{{\rm{BS1}}}^{'} = \frac{{{\phi _{{\rm{th}}}}{P_i}L_{ik}^S}}{{G_{n{\text{U}}_{{\rm{N}}2}^j}^tG_{n{\text{U}}_{{\rm{N}}2}^j}^rL_{mj}^S}}$.

\subsection{BS Outage Probability Analysis}

Meanwhile, using (6) and (12), the OP of the ${\rm{BS}}_i$-${{\rm{U}}_{\rm{BS}}^t}$ link in the proposed JMDR-IM scheme is written as
\begin{eqnarray}
&&\!\!\!\!\!\!\!\!\!\!\!\!\!{P_{\rm{OP}}^{\rm{BS}}}= \Pr\Bigg(\nonumber\\
&&\!\!\!\!\!\!\!\!\!\!\!\!\!\!\frac{{{P_{i}^o}L_{it}^S}{{\left| {{h_{it}}} \right|}^2}}{{\sum\limits_{n = 1}^{{N_{{\rm{S3}}}^{'}}}\!\!{{P_{n}^{o}}G_{n{{\rm{U}}_{\rm{BS}}^t}}^t\!\!\!L_{nt}^S}{{\left|\! {{h_{nt}}} \!\right|}^2} \!\!+\!\!\!\!\! \sum\limits_{m = 1}^{{N_{{\rm{S4}}}^{'}}}\!\!{{P_{m}^{o}}L_{mt}^S}G_{m{{\rm{U}}_{\rm{BS}}^t}}^t{{\left|\! {{h_{mt}}} \!\right|}^2}\!\!\!\!\!+\!\!\!\!\!\!\!\!\!\sum\limits_{j \in \Phi _{\rm{BS}}, j \ne i}\!\!\!\!\!\!\!\!\!\!\!{{P_{j}^o}\!L_{jt}^S}\!{{\left| {{h_{jt}}} \right|}^2} \!\!+\!\! \sigma _n^2}}\nonumber\\
&& <{\phi _{th}}\Bigg),\label{equ:20}
\end{eqnarray}
where we have $L_{it}^S=L_{i{{\rm{U}}_{\rm{BS}}^t}}^{-\alpha _t}$, $L_{jt}^S=L_{j{{\rm{U}}_{\rm{BS}}^t}}^{-\alpha _t}$, with $L_{i{{\rm{U}}_{\rm{BS}}^t}}$ and $L_{j{{\rm{U}}_{\rm{BS}}^t}}$ representing the distance between ${\rm{BS}}_i$ and ${\rm{U}}_{\rm{BS}}^t$, as well as between ${\rm{BS}}_j$ and ${\rm{U}}_{\rm{BS}}^t$.

Using (A.1)-(A.3), ${P_{\rm{OP}}^{\rm{BS}}}$ can be obtained as
\begin{eqnarray}
&&\!\!\!\!\!\!\!\!\!\!\!\!\!\!{P_{\rm{OP}}^{\rm{BS}}}= \prod\limits_{n \!=\! 1}^{N_{\rm{S3}}^{'}} {\sum\limits_{{n_{{{\rm{N1}}_n}}} \!=\! 0}^{{m_{{{\rm{N1}}_n}}} \!-\! 1}\!  } {Z_{{{\rm{N1}}_n}}}\prod\limits_{m \!=\! 1}^{N_{\rm{S4}}^{'}} \!{\sum\limits_{{n_{{{\rm{N2}}_m}}} \!=\! 0}^{{m_{{{\rm{N2}}_m}}} \!-\! 1}  } {Z_{{{\rm{N2}}_m}}}-\nonumber \\
&&\;\;\;\prod\limits_{n \!=\! 1}^{N_{\rm{S3}}^{'}}\! {\sum\limits_{{n_{{{\rm{N1}}_n}}} \!=\! 0}^{{m_{{{\rm{N1}}_n}}} \!\!-\!\! 1}\!\!\!\! {Z_{{{\rm{N1}}_n}}}\left({\beta _{{{\rm{N1}}_n}}}\!\!\! -\! {\delta _{{{\rm{N1}}_n}}}\right) }\! {\left(\!\! {{\beta _{{{\rm{N1}}_n}}} \!\!\!\!-\!\! {\delta _{{{\rm{N1}}_n}}} \!+\! \frac{{{A_n}}}{{\sigma _{\rm{BS}}^2}}} \!\!\right)^{ -\! {n_{{{\rm{N1}}_n}}}\! -\! 1}}\nonumber \\
&&\;\;\;\prod\limits_{m \!=\! 1}^{N_{\rm{S4}}^{'}} \!{\sum\limits_{{n_{{{\rm{N2}}_m}}} \!=\! 0}^{{m_{{{\rm{N2}}_m}}} \!-\! 1}\!\!\!\!\! {Z_{{{\rm{N2}}_m}}}\!\!\!\left({\beta _{{{\rm{N2}}_m}}} \!\!\!\!-\! {\delta _{{{\rm{N2}}_m}}}\right) }\! {\left(\!\! {{\beta _{{{\rm{N2}}_m}}}\!\!\!\! -\! {\delta _{{{\rm{N2}}_m}}} \!\!\!+\! \frac{{{B_m}}}{{\sigma _{\rm{BS}}^2}}} \!\!\right)^{ -\!\! {n_{{{\rm{N2}}_m}}} \!\!-\!\! 1}}\nonumber \\
&&\;\;\;\;\;\;\;\;\;\;\;\;\;\;\;\;\;\;\prod\limits_{j = 1,j \ne i}^{{N_{\rm{BS}}}} {\left( {\frac{1}{{\sigma _{\rm{BS}}^2{C_j} + 1}}} \right) \cdot {e^{ - \frac{1}{{\sigma _{\rm{BS}}^2}}{D_i}}}},\label{equ:21}
\end{eqnarray}

where we have ${A_n} = \frac{{{\phi _{th}}P_n^oG_{nU_{\rm{BS}}^t}^tL_{nk}^S}}{{{P_i}L_{ik}^S}}$, ${B_m} = \frac{{{\phi _{th}}P_m^oG_{mU_{\rm{BS}}^t}^tL_{mt}^S}}{{{P_i}L_{ik}^S}}$, ${C_j} = \frac{{{\phi _{th}}P_j^oL_{jt}^S}}{{{P_i}L_{ik}^S}}$ and ${{\text{D}}_i} = \frac{{{\phi _{th}}\sigma _n^2}}{{{P_i}L_{ik}^S}}$.

When $P_i$ becomes very large, using (B.3), the AOP of ${\rm{BS}}_i$-${{\rm{U}}_{\rm{BS}}^t}$ link in the proposed JMDR-IM scheme can also be obtained.

\section{Simulation Results}
In this section, we evaluate the proposed JMDR-IM scheme in terms of its SINR characteristics and the OP of the SGIN considered, and compare it to the conventional PPAFB scheme. Moreover, the analytic OPs of NGSO 1, of NGSO 2 and of the BSs are evaluated by plotting (A.7), (18) and (21), respectively. The simulated OPs of NGSO 1, of NGSO 2 and of the BSs are also provided to verify the accuracy of the the analytic OPs, which are denoted by (S.). Furthermore, the analytic AOPs of NGSO 1, of NGSO 2 and of the BSs, denoted by (A.), are analyzed by plotting (16), (19) and (B.3). The associated simulation parameters are given in Table III. Moreover, the experimental platform is an Intel(R) Core(TM)i7-9700K CPU 3.60GHz, the GPU is NVIDIA RTX2080Ti, and the amount of memory is 64.00GB. The neural network is constructed on the Tensorflow framework. Note that the default number of NGSO constellations is two in the simulations, and these two constellations are NGSO 1 and NGSO 2. Moreover, the size of the dataset employed for both training and testing the LSTM are ten thousand and one-hundred thousand.

\begin{table}[htbp]
\scriptsize
\renewcommand\arraystretch{1}
\centering
\caption{Simulation Parameters}
\begin{tabular}{l|c|c}
\bottomrule
Parameter&Symbol&Value\\
\hline
NGSO 1 satellite number & $N$&648\\
\hline
NGSO 2 satellite number&$M$&1584\\
\hline
NGSO 3 satellite number& &720\\
\hline
NGSO 1 satellite antenna diameter& $D_{\rm{N1}_n}$&0.4 m\\
\hline
NGSO 2 satellite antenna diameter& $D_{\rm{N2}_m}$&0.4 m\\
\hline
NGSO 3 satellite antenna diameter& &0.4 m\\
\hline
NGSO 1 user antenna diameter& $D_{{\rm{U}}_{{\rm{N1}}_k}}$&0.6 m\\
\hline
NGSO 2 user antenna diameter& $D_{{\rm{U}}_{{\rm{N2}}_j}}$&0.6 m\\
\hline
NGSO 3 user antenna diameter& &0.6 m\\
\hline
Antenna efficiency & $\xi $&$55\%$\\
\hline
Frequency band & $f$&17.9 GHz\\
\hline
NGSO 1 transmission bandwidth & $B_n $&125 MHz\\
\hline
NGSO 2 transmission bandwidth & $B_m $&125 MHz\\
\hline
NGSO 3 transmission bandwidth & &125 MHz\\
\hline
NGSO 1 orbit altitude& $h_{\rm{N1}_n}$&500 km\\
\hline
NGSO 2 orbit altitude& $h_{\rm{N2}_m}$&550 km\\
\hline
NGSO 3 orbit altitude& &570 km\\
\hline
Earth radius& $R$&6370 km\\
\hline
Equivalent noise temperature& $T_n$&290 K\\
\hline
NGSO 1 maximum transmit power& $P_{\rm{NGSO1}}^{\max }$&5 W\\
\hline
NGSO 2 maximum transmit power& $P_{\rm{NGSO2}}^{\max }$&5 W\\
\hline
NGSO 3 maximum transmit power& &5 W\\
\hline
BS maximum transmit power& $P_{\rm{BS}}^{\max }$&5 W\\
%\hline
%HPPP density&$\lambda _{\rm{BS}}$&$1.6\!\!\times\!\! 10^{\!-4}\!\!/{{\rm{km}}^2}$\\
\hline
SINR threshold& $\phi _{th}$&10 dB\\
\hline
Greedy probability& $\varepsilon$-greedy&0.8\\
\hline
Scattered components's average power& $c_{{\rm{N1}}_n}$($c_{{\rm{N2}}_m}$)&0.126\\
\hline
Nakagami fading parameters& $m_{{\rm{N1}}_n}$($m_{{\rm{N2}}_m}$)&10.1\\
\hline
LOS components's average power& $\Omega _{{\rm{N1}}_n}$($\Omega _{{\rm{N2}}_m}$)&0.835\\
\hline
Channel-gain mean& ${\sigma _{\rm{BS}}^{2}}$&1\\
\toprule
\end{tabular}
\end{table}

Figure~\ref{fig:fig7} portrays ${\rm{SINR}}_{{\rm{U}}_{\rm{N1}}^k}$ versus $P_{\rm{NGSO1}}^{\max }$ parameterized by different $\theta ^{\rm{BW}}$ for the proposed JMDR-IM and PPAFB schemes. Observe from Fig.~\ref{fig:fig7} that the proposed JMDR-IM scheme outperforms the conventional PPAFB scheme in terms of its ${\rm{SINR}}_{{\rm{U}}_{\rm{N1}}^k}$ for both $\theta ^{\rm{BW}} = 15^{o}$ and $\theta ^{\rm{BW}} = 7.5^{o}$. Furthermore, the proposed JMDR-IM scheme of $\theta ^{\rm{BW}} = 15^{o}$ achieves a higher ${\rm{SINR}}_{{\rm{U}}_{\rm{N1}}^k}$ than the PPAFB scheme of $\theta ^{\rm{BW}} = 7.5^{o}$. It is also shown in Fig.~\ref{fig:fig7} that the ${\rm{SINR}}_{{\rm{U}}_{\rm{N1}}^k}$ of the JMDR-IM scheme for $\theta ^{\rm{BW}} = 15^{o}$ and of the PPAFB scheme for both $\theta ^{\rm{BW}} = 15^{o}$ and $\theta ^{\rm{BW}} = 7.5^{o}$ tends to be constant. This near-constant ${\rm{SINR}}_{{\rm{U}}_{\rm{N1}}^k}$ is achieved, since increasing the transmit power of NGSO 1 may increase the interference imposed on the users of NGSO 2 and of the BSs. To meet the transmission requirements of both NGSO 2 and of the BSs, the NGSO 1 satellites are not allowed to increase their transmit power beyond a certain threshold. By contrast, the JMDR-IM scheme can have a higher transmit power, demonstrating the superiority of the JMDR-IM scheme in terms of managing interference. Observe from Fig.~\ref{fig:fig7} that the SINR in the JMDR-IM and the PPAFB schemes are reduced in the presence of the intra-constellation CFI, compared to no intra-constellation CFI. This SINR reduction may be tolerable, especially when more robust multi-color multibeam frequency-reuse schemes are used, such as the FR7 multibeam \cite{BIBFR7}, the eight-color frequency-reuse \cite{BIBFR8}, and the twelve-color frequency-reuse multibeam regimes \cite{BIBFR12}. Additionally, as shown in Fig.~\ref{fig:fig7}, the SINR of both the JMDR-IM and the PPAFB schemes using the predicted CSI, as well as that of the JMDR-IM and the PPAFB schemes was quantified by using the actual CSI, showing the effectiveness of the LSTM-ARMA-DQN conceived in terms of predicating the CSI. We also analyze the case when the intra-constellation interference is considered, but the inter-constellation interference is neglected. One can observe from Fig.~\ref{fig:fig7} that this case achieves the highest SINR, albeit at the cost of requiring more spectral resources. This result demonstrates that the effect on the SINR imposed by the inter-constellation CFI is more severe than that inflicted by the intra-constellation CFI in the SGIN considered.

\begin{figure}[htbp]
\centering
\includegraphics[width=0.95\linewidth]{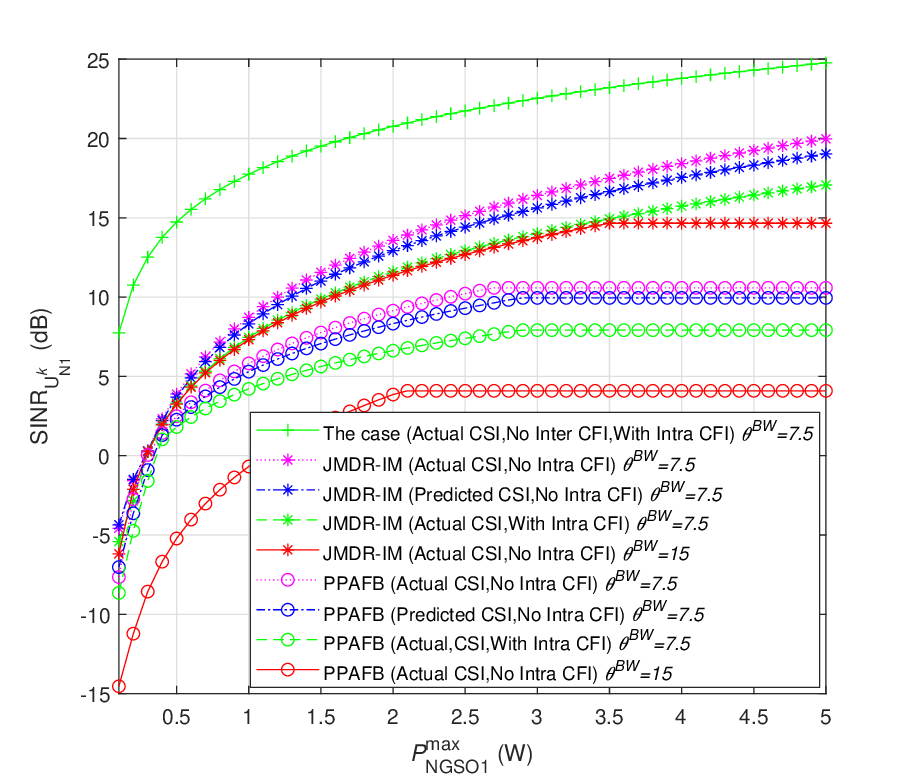}
\caption{${\rm{SINR}}_{{\rm{U}}_{\rm{N1}}^k}$ versus $P_{\rm{NGSO1}}^{\max }$ parameterized by different $\theta ^{\rm{BW}}$ for the proposed JMDR-IM and the conventional PPAFB schemes, wherein the ${\rm{SINR}}_{{\rm{U}}_{\rm{N1}}^k}$ is evaluated by optimizing (11), the effect of the intra-constellation CFI, the actual CSI and the predicted CSI are investigated. \label{fig:fig7}}
\end{figure}

Figure~\ref{fig:fig8} depicts the ${\rm{SINR}}_{{\rm{U}}_{\rm{N2}}}^{\rm{min}}$ versus $P_{\rm{NGSO1}}^{\max }$ parameterized by different $\theta ^{\rm{BW}}$ for the proposed JMDR-IM and the conventional PPAFB schemes, where ${\rm{SINR}}_{{\rm{U}}_{\rm{N2}}}^{\rm{min}}$ denotes the lowest SINR of different ${{\rm{U}}_{\rm{N2}}^j}$. It is shown from Fig.~\ref{fig:fig8} that the ${\rm{SINR}}_{{\rm{U}}_{\rm{N2}}}^{\rm{min}}$ of the proposed JMDR-IM scheme is higher than that of the conventional PPAFB scheme, when $P_{\rm{NGSO1}}^{\max }$ is lower than 3.5 W. Fig.~\ref{fig:fig8} also shows that although a higher transmit power is available, NGSO 1 relying on the conventional PPAFB scheme may perform wireless transmissions at a limited transmit power for satisfying the transmission requirements of both NGSO 2 and of the BSs. However, the proposed JMDR-IM scheme allows NGSO 1 to perform wireless transmissions at a higher transmit power than the traditional PPAFB scheme, whilst meeting the transmission requirements of both NGSO 2 and of the BSs. This increased transmit power of NGSO 1 can be effectively converted into a SINR improvement of NGSO 1, as shown in Fig.~\ref{fig:fig7}.

\begin{figure}[htbp]
\centering
\includegraphics[width=1\linewidth]{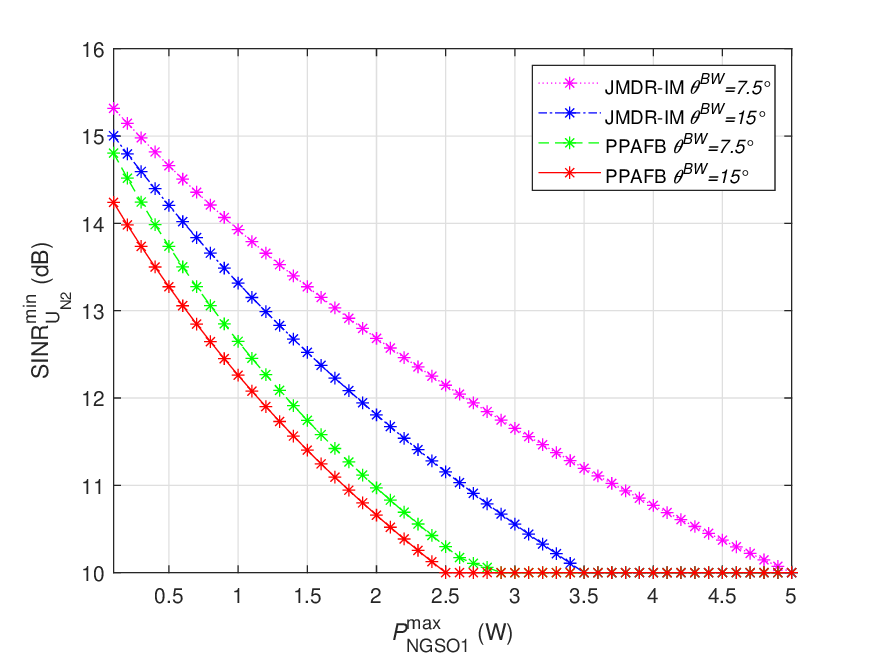}
\caption{${\rm{SINR}}_{{\rm{U}}_{\rm{N2}}}^{\rm{min}}$ versus $P_{\rm{NGSO1}}^{\max }$ parameterized by different $\theta ^{\rm{BW}}$ for the proposed JMDR-IM and the conventional PPAFB schemes, wherein NGSO 1 and NGSO 2 constellations are considered.\label{fig:fig8}}
\end{figure}

Figure~\ref{fig:fig9} explores the ${\rm{SINR}}_{{\rm{U}}_{\rm{N2}}}^{\rm{min}}$ versus $P_{\rm{NGSO2}}^{\max }$ parameterized by different number of NGSO constellations for both the proposed JMDR-IM and for the conventional PPAFB schemes. Observe from Fig.~\ref{fig:fig9} that although NGSO 2 relying on both the JMDR-IM and the PPAFB schemes meets the SINR threshold, the required transmit power of NGSO 2 using the JMDR-IM scheme is lower than that of NGSO 2 employing the PPAFB scheme. This is due to the fact that given an SINR threshold, a lower transmit power is required in the presence of lower interference received. These beneficial results indicate that the JMDR-IM scheme effectively manages the interference imposed on the users of NGSO 2. It is also shown from Fig.~\ref{fig:fig9} that as the number of NGSO constellations increases from two to three, the SINR of users in NGSO 2 decreases correspondingly, when using the JMDR-IM and the PPAFB schemes. This result shows that the inter-constellation CFI becomes more severe upon increasing the number of spectrum-sharing NGSO constellations. Fortunately, the proposed JMDR-IM scheme can still meet the transmission requirements even in the presence of three NGSO constellations.

\begin{figure}[htbp]
\centering
\includegraphics[width=1\linewidth]{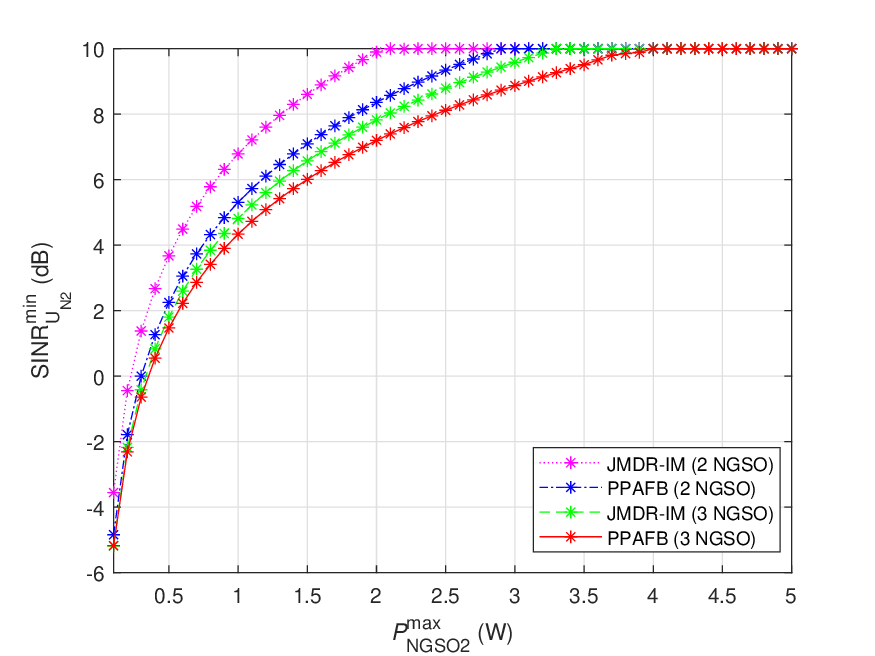}
\caption{${\rm{SINR}}_{{\rm{U}}_{\rm{N2}}}^{\rm{min}}$ versus $P_{\rm{NGSO2}}^{\max }$ parameterized by different number of NGSO constellations for the proposed JMDR-IM and the conventional PPAFB schemes, wherein $\theta ^{\rm{BW}} = 7.5^{o}$.\label{fig:fig9}}
\end{figure}

Figure~\ref{fig:fig10} shows the ${\rm{SINR}}_{{\rm{U}}_{\rm{N1}}^k}$ versus $P_{\rm{NGSO2}}^{\max }$ parameterized by different number of NGSO constellations for both the proposed JMDR-IM, the conventional PPAFB and the the pure fixed power based fixed beams (PFPFB) schemes. Similarly to Fig.~\ref{fig:fig9}, one can observe from Fig.~\ref{fig:fig10} that although a higher $P_{\rm{NGSO2}}^{\max }$ is available, the proposed JMDR-IM scheme allocates a lower transmit power than the PPAFB scheme for NGSO 2. This lower transmit power of NGSO 2 using the JMDR-IM scheme not only readily meets the SINR requirements of NGSO 2, but also reduces the amount of interference imposed on the users of NGSO 1, hence resulting in a higher ${\rm{SINR}}_{{\rm{U}}_{\rm{N1}}^k}$. Fig.~\ref{fig:fig10} also shows that the SINR will be reduced in the presence of three NGSO constellations. However, the proposed JMDR-IM scheme still achieves a higher SINR than the PPAFB and the PFPFB schemes, when the number of spectrum-sharing NGSO constellations is three. Compared to the PPAFB scheme, the performance gain of beam shut-off/beam-switching can be evaluated. This result demonstrates the superiority of the beam shut-off/beam-switching regime in terms of managing the inter-constellation CFI. Moreover, this gain increases upon increasing $P_{\rm{NGSO2}}^{\max }$. These beneficial results indicate that the CFI management of beam shut-off/beam-switching and deep-learning-based power allocation harnessed in the proposed JMDR-IM scheme can be effectively converted into SINR enhancements.

\begin{figure}[htbp]
\centering
\includegraphics[width=1\linewidth]{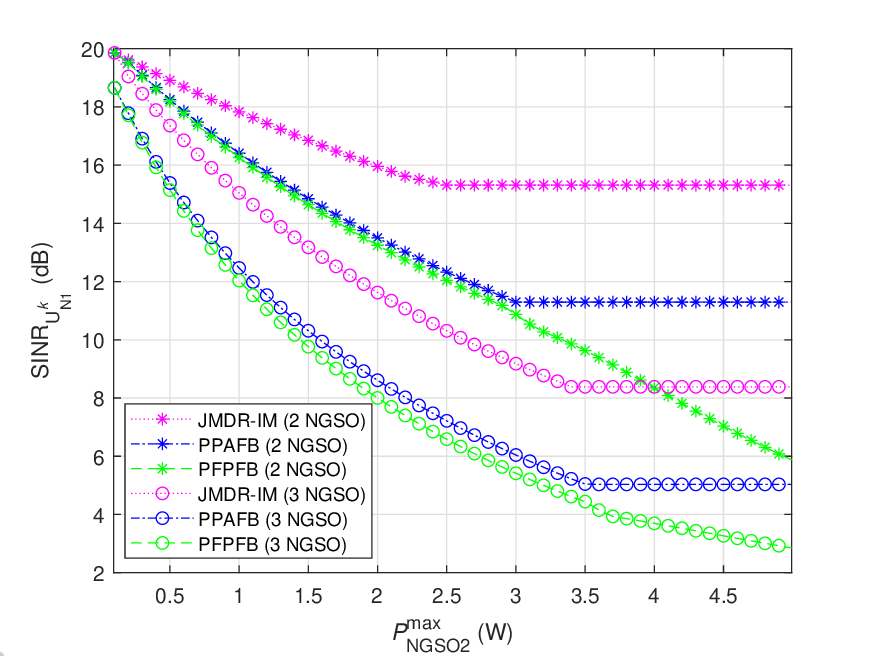}
\caption{${\rm{SINR}}_{{\rm{U}}_{\rm{N1}}^k}$ versus $P_{\rm{NGSO2}}^{\max }$ parameterized by different number of NGSO constellations for the proposed JMDR-IM, the conventional PPAFB and the PFPFB schemes, wherein $\theta ^{\rm{BW}} = 7.5^{o}$.\label{fig:fig10}}
\end{figure}

In Fig.~\ref{fig:fig11}, we investigate $C/(I+N)$ of NGSO 1 at different time instants. Observe from Fig.~\ref{fig:fig2} that NGSO 2 may impose a higher amount of interference on the users of NGSO 1 in the presence of spectrum sharing without using the JMDR-IM scheme. Moreover, the interference imposed by NGSO 2 may fluctuate as the satellites of NGSO 2 move. These interference sources may result in a low and time-varying $C/(I+N)$ for NGSO 1. In contrast to Fig.~\ref{fig:fig2}, observe from Fig.~\ref{fig:fig11} that the interference imposed by NGSO 2 is beneficially managed by the proposed JMDR-IM scheme. Additionally, the JMDR-IM scheme can track the time-varying interference, and maintain a high $C/(I+N)$ for NGSO 1. These beneficial results demonstrate the superiority of the proposed JMDR-IM scheme in a spectrum-sharing SGIN in the face of dynamically fluctuating propagation characteristics.

\begin{figure}[htbp]
\centering
\includegraphics[width=1\linewidth]{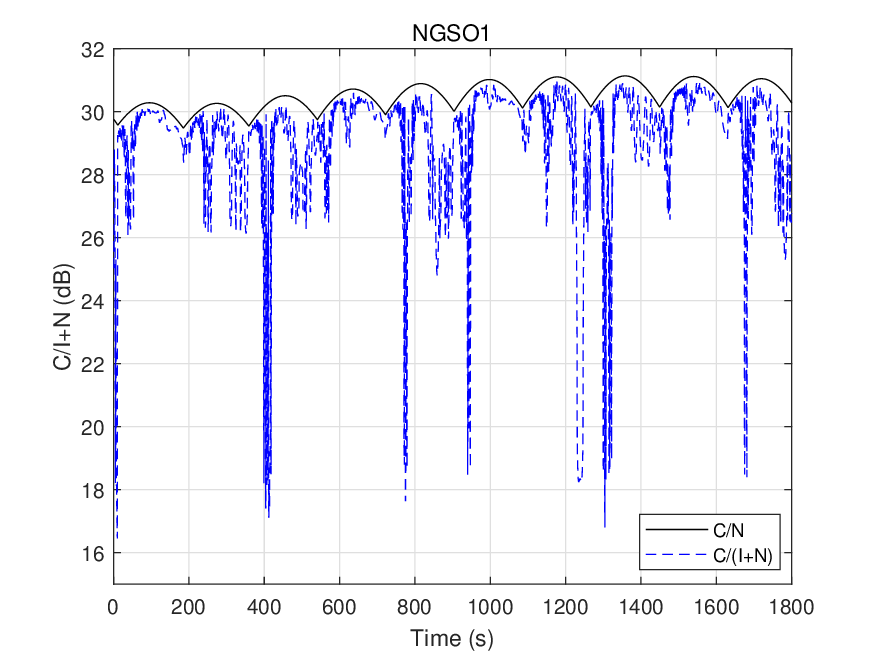}
\caption{$C/(I+N)$ of NGSO 1 versus time using the proposed JMDR-IM scheme.\label{fig:fig11}}
\end{figure}

Figure~\ref{fig:fig12} explores ${P_{\rm{OP}}^{\rm{NGSO1}}}$ versus the SINR threshold $\phi _{th}$ parameterized by different $\theta ^{\rm{BW}}$ for the proposed JMDR-IM and the conventional PPAFB schemes, wherein ${P_{\rm{AOP}}^{\rm{NGSO1}}}$ is also evaluated. Observe from Fig.~\ref{fig:fig12} that the proposed JMDR-IM scheme outperforms the conventional PPAFB scheme in terms of its OP for both $\theta ^{\rm{BW}} = 15^{o}$ and $\theta ^{\rm{BW}} = 7.5^{o}$. Furthermore, given $\phi _{th} = 10$ dB, ${P_{\rm{AOP}}^{\rm{NGSO1}}}$ approaches $10^{-3}$. These beneficial results demonstrate the superiority of the JMDR-IM scheme in terms of increasing the reliability of wireless transmissions in the SGIN considered. Fig.~\ref{fig:fig12} also shows that ${P_{\rm{OP}}^{\rm{NGSO1}}}$ is close to ${P_{\rm{AOP}}^{\rm{NGSO1}}}$ in the low $\phi _{th}$ region, and ${P_{\rm{OP}}^{\rm{NGSO1}}}$ is a bit lower than the ${P_{\rm{AOP}}^{\rm{NGSO1}}}$ in the high $\phi _{th}$ region. Moreover, the analytic ${P_{\rm{OP}}^{\rm{NGSO1}}}$ matches the simulated ${P_{\rm{OP}}^{\rm{NGSO1}}}$, verifying the accuracy of the analysis.

\begin{figure}[htbp]
\centering
\includegraphics[width=1\linewidth]{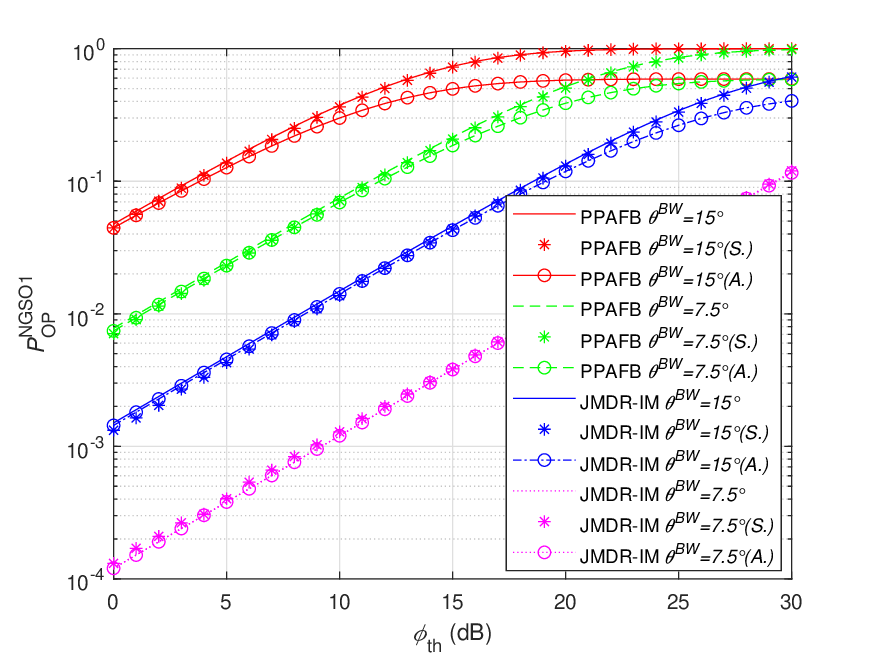}
\caption{${P_{\rm{OP}}^{\rm{NGSO1}}}$ versus the SINR threshold $\phi _{th}$ parameterized by different $\theta ^{\rm{BW}}$ for the proposed JMDR-IM and the conventional PPAFB schemes, wherein the analytic ${P_{\rm{OP}}^{\rm{NGSO1}}}$ and ${P_{\rm{AOP}}^{\rm{NGSO1}}}$ are evaluated by plotting (A.7) and (16), respectively.\label{fig:fig12}}
\end{figure}

Figure~\ref{fig:fig13} investigates ${P_{\rm{OP}}^{\rm{NGSO2}}}$ versus the SINR threshold $\phi _{th}$ parameterized by different $\theta ^{\rm{BW}}$ for the proposed JMDR-IM and the conventional PPAFB schemes, wherein ${P_{\rm{AOP}}^{\rm{NGSO2}}}$ is also evaluated. Observe from Fig.~\ref{fig:fig13} that the proposed JMDR-IM scheme achieves a lower ${P_{\rm{OP}}^{\rm{NGSO2}}}$ than the conventional PPAFB scheme. Furthermore, observe from Fig.~\ref{fig:fig13} that given a tolerable OP, the $\phi _{th}$ of the proposed JMDR-IM scheme is higher than that of the conventional PPAFB scheme. This is due to the fact that the proposed JMDR-IM scheme effectively manages the interference received at NGSO 2 and attains a higher SINR than the PPAFB scheme.

\begin{figure}[htbp]
\centering
\includegraphics[width=1\linewidth]{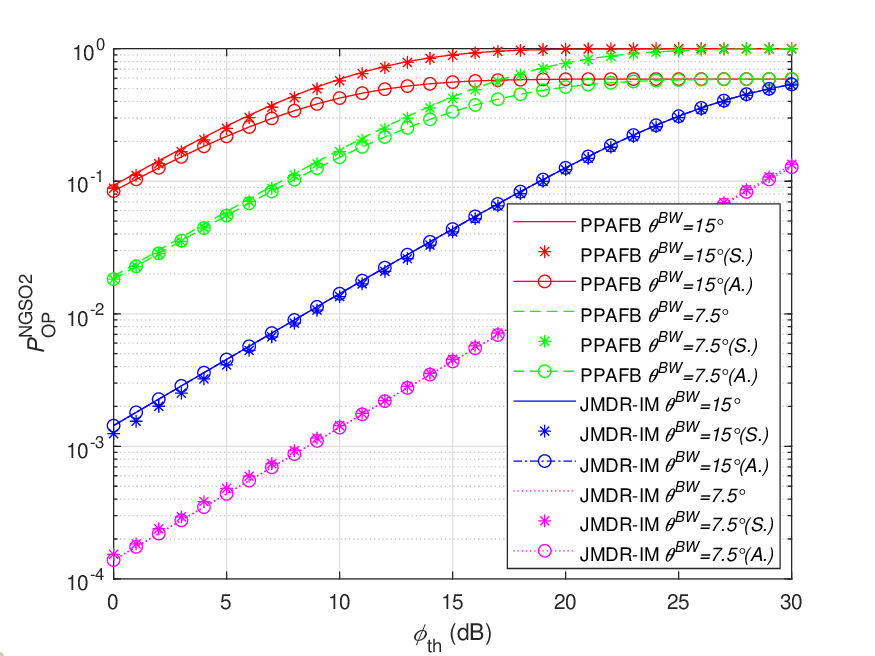}
\caption{${P_{\rm{OP}}^{\rm{NGSO2}}}$ versus the SINR threshold $\phi _{th}$ parameterized by different $\theta ^{\rm{BW}}$ for the proposed JMDR-IM and the conventional PPAFB schemes, wherein the analytic ${P_{\rm{OP}}^{\rm{NGSO2}}}$ and ${P_{\rm{AOP}}^{\rm{NGSO2}}}$ are evaluated by plotting (18) and (19), respectively.\label{fig:fig13}}
\end{figure}

In Fig.~\ref{fig:fig14}, we analyze ${P_{\rm{OP}}^{\rm{BS}}}$ versus the SINR threshold $\phi _{th}$ parameterized by different $\theta ^{\rm{BW}}$ for the proposed JMDR-IM and the conventional PPAFB schemes, wherein ${P_{\rm{AOP}}^{\rm{BS}}}$ is evaluated. As shown in Fig.~\ref{fig:fig14}, when the $\phi _{th}$ is reduced from 30 dB to 0 dB, the ${P_{\rm{OP}}^{\rm{BS}}}$ of both the JMDR-IM and PPAFB schemes decreases correspondingly. Moreover, observe from Fig.~\ref{fig:fig14} that for a specific $\phi _{th}$, the proposed JMDR-IM scheme still achieves a lower OP for the BSs than the conventional PPAFB scheme, through managing the interference imposed on the users of the BSs. Additionally, the ${P_{\rm{AOP}}^{\rm{BS}}}$ equals the ${P_{\rm{OP}}^{\rm{BS}}}$ in the low $\phi _{th}$ region, and the ${P_{\rm{AOP}}^{\rm{BS}}}$ becomes lower than the the ${P_{\rm{OP}}^{\rm{BS}}}$ when the $\phi _{th}$ approaches 3 dB. Upon combining Fig.~\ref{fig:fig12}$\sim$Fig.~\ref{fig:fig14}, we can infer that the proposed JMDR-IM scheme simultaneously decreases the OPs of NGSO 1, of NGSO 2 and of the BSs. This significant OP deduction confirms the superiority of the proposed JMDR-IM scheme in terms of managing the interference and increasing the reliability of wireless transmissions.
\begin{figure}[htbp]
\centering
\includegraphics[width=1\linewidth]{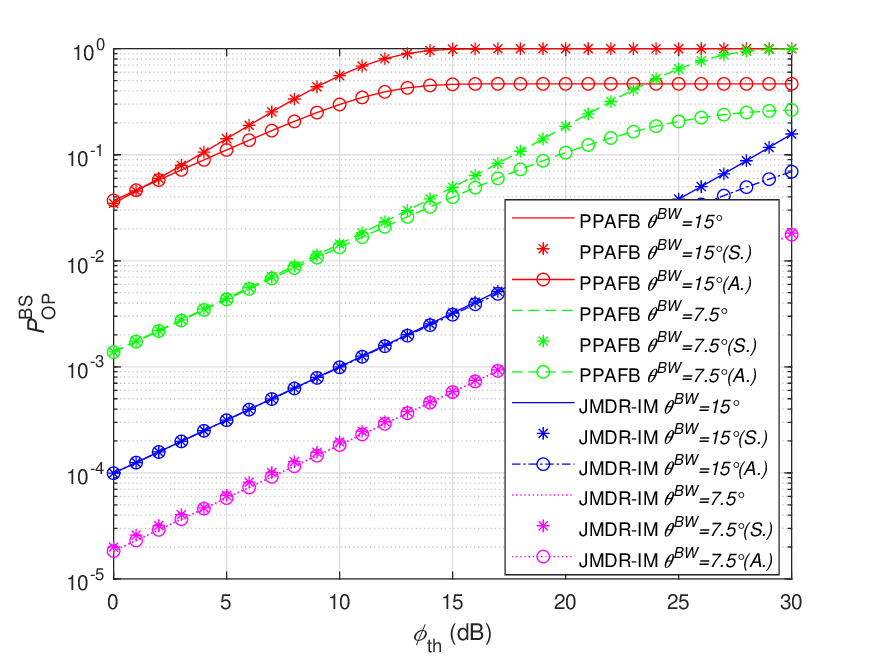}
\caption{${P_{\rm{OP}}^{\rm{BS}}}$ versus the SINR threshold $\phi _{th}$ parameterized by $\theta ^{\rm{BW}}$ for the proposed JMDR-IM and the conventional PPAFB schemes, wherein the analytic ${P_{\rm{OP}}^{\rm{BS}}}$ and ${P_{\rm{AOP}}^{\rm{BS}}}$ are evaluated by plotting (21) and (B.3).\label{fig:fig14}}
\end{figure}

\section{Conclusions}
Spectrum sharing in SGINs is capable of improving the spectrum utilization, which however leads to severe CFI. We proposed the JMDR-IM scheme for managing the CFI, relying on our coverage overlap analysis, and on multi-domain resource scheduling. To be specific, the coverage overlap of a pair of NGSO constellations was analyzed, and the co-frequency exclusion zone concept was harnessed for managing the co-line interference. Then, multi-domain resources were jointly scheduled relying on the analysis results. More specifically, beam shut-off and beam-switching aided scheduling was harnessed for alleviating the CFI. Then, the LSTM-ARMA-DQN scheme was adopted for alleviating the CFI via optimizing the transmit power for NGSO 1, of NGSO 2 and of the BSs. Additionally, the OP and AOP were also analyzed, to demonstrate the superiority of the proposed JMDR-IM scheme in terms of guaranteeing reliable transmissions in the SGIN considered.

\numberwithin{equation}{section}
\appendices
\section{Derivation of ${P_{\rm{OP}}^{\rm{NGSO1}}}$}
According to \cite{BIB29}, the probability density function (PDF) of the random variables (RVs) ${{| {{h_{nk}}} |}^2}$, ${{| {{h_{mk}}} |}^2}$ and ${{| {{h_{ik}}} |}^2}$ can be respectively formulated as:
\begin{eqnarray}
&&\!\!\!\!\!\!\!\!\!\!\!{f_{{{\left|  {{h_{nk}}}  \right|}^2}}}\left( x \right) \nonumber\\
&&\!\!\!\!\!\!\!\!\!\!\!= {\alpha _{{\rm{N1}}_n}}{e^{ - {\beta _{{\rm{N1}}_n}}x}}{}_1{F_1}\left( {{m_{{\rm{N1}}_n}};1;{\delta _{{\rm{N1}}_n}}x} \right)\nonumber \\
&&\!\!\!\!\!\!\!\!\!\!\!={\alpha _{{\rm{N1}}_n}}{e^{ - \left( {{\beta _{{\rm{N1}}_n}} - {\delta _{{\rm{N1}}_n}}} \right)x}}\nonumber \\
&&\!\!\!\sum\limits_{{n_{{\rm{N1}}_n}} = 0}^{{m_{{\rm{N1}}_n}} - 1} \!\!{\frac{{\left( {{m_{{\rm{N1}}_n}} - 1} \right)!{{\left( {{\delta _{{\rm{N1}}_n}}x} \right)}^{{n_{{\rm{N1}}_n}}}}}}{{\left( {{m_{{\rm{N1}}_n}} \!-\! 1 \!-\! {n_{{\rm{N1}}_n}}} \right)!{n_{{\rm{N1}}_n}}!{{\left( 1 \right)}_{{n_{{\rm{N1}}_n}}}}}}}
\end{eqnarray}
and
\begin{eqnarray}
&&\!\!\!\!\!\!\!\!\!\!\!{f_{{{\left| {{h_{mk}}} \right|}^2}}}\left( x \right)\nonumber\\
&&\!\!\!\!\!\!\!\!\!\!\!= {\alpha _{{\rm{N2}}_m}}{e^{ - {\beta _{{\rm{N2}}_m}}x}}{}_1{F_1}\left( {{m_{{\rm{N2}}_m}};1;{\delta _{{\rm{N2}}_m}}x} \right)\nonumber \\
&&\!\!\!\!\!\!\!\!\!\!={\alpha _{{\rm{N2}}_m}}{e^{ - \left( {{\beta _{{\rm{N2}}_m}} - {\delta _{{\rm{N2}}_m}}} \right)x}}\nonumber\\
&&\!\!\!\sum\limits_{{n_{{\rm{N2}}_m}} = 0}^{{m_{{\rm{N2}}_m}} - 1} \!\!\!{\frac{{\left( {{m_{{\rm{N2}}_m}} - 1} \right)!{{\left( {{\delta _{{\rm{N2}}_m}}x} \right)}^{{n_{{\rm{N2}}_m}}}}}}{{\left( {{m_{{\rm{N2}}_m}} \!-\! 1 \!-\! {n_{{\rm{N2}}_m}}} \right)!{n_{{\rm{N2}}_m}}!{{\left( 1 \right)}_{{n_{{\rm{N2}}_m}}}}}}}
\end{eqnarray}
and
\begin{equation}
{f_{{{\left| {{h_{ik}}} \right|}^2}}}\left( x \right) = \frac{1}{\sigma _{\rm{BS}}^{2}}{e^{ - \frac{x}{{\sigma _{\rm{BS}}^{2}}}}},
\end{equation}
where ${}_1{F_1}( { \cdot , \cdot , \cdot } )$ is the confluent hypergeometric function, ${\alpha _{{\rm{N2}}_m}} = \frac{{{{( {2{c_{{{\rm{N2}}_m}}}{m_{{{\rm{N2}}_m}}}} )}^{{m_{{{\rm{N2}}_m}}}}}}}{{2{c_{{{\rm{N2}}_m}}}{{( {2{c_{{{\rm{N2}}_m}}}{m_{{{\rm{N2}}_m}}} + {\Omega _{{{\rm{N2}}_m}}}} )}^{m_{{\rm{N2}}_m}}}}}$, ${\alpha _{{\rm{N1}}_n}} = \frac{{{{( {2{c_{{{\rm{N1}}_n}}}{m_{{{\rm{N1}}_n}}}} )}^{{m_{{{\rm{N1}}_n}}}}}}}{{2{c_{{{\rm{N1}}_n}}}{{( {2{c_{{{\rm{N1}}_n}}}{m_{{{\rm{N1}}_n}}} + {\Omega _{{{\rm{N1}}_n}}}} )}^{m_{{\rm{N1}}_n}}}}}$, ${\delta _{{{\rm{N1}}_n}}} = \frac{{{\Omega _{{{\rm{N1}}_n}}}}}{{2{c_{{{\rm{N1}}_n}}}( {2{c_{{{\rm{N1}}_n}}}{m_{{{\rm{N1}}_n}}} + {\Omega _{{{\rm{N1}}_n}}}} )}}$, ${\delta _{{{\rm{N2}}_m}}} = \frac{{{\Omega _{{{\rm{N2}}_m}}}}}{{2{c_{{{\rm{N2}}_m}}}( {2{c_{{{\rm{N2}}_m}}}{m_{{{\rm{N2}}_m}}} + {\Omega _{{{\rm{N2}}_m}}}} )}}$, ${\beta _{{{\rm{N1}}_n}}} = \frac{1}{{2{c_{{{\rm{N1}}_n}}}}}$ and ${\beta _{{{\rm{N2}}_m}}} = \frac{1}{{2{c_{{{\rm{N2}}_m}}}}}$. $2{c_{{{\rm{N1}}_n}}}$ and $2{c_{{{\rm{N2}}_m}}}$ denote the scattered components' average power of the ${\rm{N1}}_n$-${{\rm{U}}_{\rm{N1}}^k}$ and of the ${\rm{N2}}_m$-${{\rm{U}}_{\rm{N2}}^j}$ links, respectively. Furthermore, ${\Omega _{{{\rm{N1}}_n}}}$ and ${\Omega _{{{\rm{N2}}_m}}}$ are the line-of-sight (LOS) components' average power of the ${\rm{N1}}_n$-${{\rm{U}}_{\rm{N1}}^k}$ and ${\rm{N2}}_m$-${{\rm{U}}_{\rm{N2}}^j}$ links, respectively, while ${\sigma _{\rm{BS}}^{2}}$ is the mean of ${{| {{h_{ik}}} |}^2}$. Additionally, ${m_{{{\rm{N1}}_n}}}$ and ${m_{{{\rm{N2}}_m}}}$ represent the Nakagami fading parameters of the ${\rm{N1}}_n$-${{\rm{U}}_{\rm{N1}}^k}$ and ${\rm{N2}}_m$-${{\rm{U}}_{\rm{N2}}^j}$ links, respectively.

Using (A.1), $P_{{\rm{OP}}}^{{\rm{NGSO1}}}$ can be written as
\begin{eqnarray}
&&\!\!\!\!\!\!\!\!\!\!\!\!\!\!\!\!\!\!\!\!\!\!\!\!P_{{\rm{OP}}}^{{\rm{NGSO1}}}\nonumber\\
&&\!\!\!\!\!\!\!\!\!\!\!\!\!\!\!\!\!\!\!\!\!\!\!\!=\!\!\Pr\!\! \left( \!\! {\frac{{P_n^oG_{n{\text{U}}_{{\rm{N1}}}^k}^tG_{n{\text{U}}_{{\rm{N1}}}^k}^rL_{nk}^S{{\left| {{h_{nk}}} \right|}^2}}}{{\sum\limits_{m = 1}^{N_{{\rm{S1}}}^{'}}\!\!\! {P_m^oG_{m{\text{U}}_{{\rm{N1}}}^k}^t\!\!G_{m{\text{U}}_{{\rm{N1}}}^k}^r\!\!L_{mk}^S{{\left|\! {{h_{mk}}} \!\right|}^2}} \!\! +\!\!\!\! \sum\limits_{i \in {\Phi _{{\text{BS}}}}} \!\!\!\!\!{P_i^oL_{ik}^S{{\left|\! {h_{ik}^t} \!\right|}^2}}  \!\!+\!\! \sigma _{{\rm{N1}}}^2}}}\right.\nonumber\\
&& \left.{< {\phi _{th}}} \right)\nonumber\\
&&\!\!\!\!\!\!\!\!\!\!\!\!\!\!\!\!\!\!\!\!\!\!\!\!=\!\! \Pr\!\! \left(\!\! {|{h_{nk}}|^2 \!\!<\!\! \sum\limits_{m = 1}^{N_{{\rm{S1}}}^{'}}\!\!\! {{\zeta _m}{{\left| {{h_{mk}}} \right|}^2} \!\!+\!\!\!\! {{\sum\limits_{i \in {\Phi _{{\text{BS}}}}} \!\!\!\!{{I_{{\rm{BS}}_i}}|{h_{ik}}|} }^2} \!\!+\!\! S\sigma _{{\rm{N1}}}^2} } \!\!\right),
\end{eqnarray}
where ${\zeta _m} = \frac{{{\phi _{\rm{th}}}P_m^oG_{m{\text{U}}_{{\rm{N1}}}^k}^tG_{m{\text{U}}_{{\rm{N1}}}^k}^rL_{mk}^s}}{{P_n^oG_{n{\text{U}}_{{\rm{N1}}}^k}^tG_{n{\text{U}}_{{\rm{N1}}}^k}^rL_{nk}^s}}$, ${I_{{\rm{BS}}_i}} = \frac{{{\phi _{\rm{th}}}P_i^oL_{ik}^s}}{{P_n^oG_{n{\text{U}}_{{\rm{N1}}}^k}^tG_{n{\text{U}}_{{\rm{N1}}}^k}^rL_{nk}^s}}$, and $S = \frac{{{\phi _{\rm{th}}}}}{{P_n^oG_{n{\text{U}}_{{\rm{N1}}}^k}^tG_{n{\text{U}}_{{\rm{N1}}}^k}^rL_{nk}^s}}$.

Upon denoting $\mu  =\!\! \sum\limits_{m = 1}^{N_{{\rm{S1}}}^{'}} {{\zeta _m}{{\left| {{h_{mk}}} \right|}^2} +\!\!\!\! {{\sum\limits_{i \in {\phi _{BS}}}\!\!\!\! {{I_{{\rm{BS}}_i}}|{h_{ik}}|} }^2} + S\sigma _{{\rm{N1}}}^2} $, and $x = |{h_{nk}}|^2$, $P_{{\rm{OP}}}^{{\rm{NGSO1}}}$ can be formulated as
\begin{eqnarray}
&&\!\!\!\!\!\!\!\!\!\!\!\!\!\!\!P_{{\rm{OP}}}^{{\rm{NGSO1}}}\nonumber\\
&&\!\!\!\!\!\!\!\!\!\!\!\!\!\!\!= \!\! \int_0^\infty  \!\!\!\!\!{\!\cdots \! \int_0^\infty\!\!\!\!\!  {f\!\!\left(\!\! {{{\left| {{h_{1k}}} \right|}^2}} \!\!\right)\!\cdots\!} } f\!\!\left(\!\! {{{\left| {{h_{N_{{S_1}}^{'}k}}} \right|}^2}} \!\!\right)d{\left|\! {{h_{1k}}} \!\right|^2} \!\cdots\! d{\left|\! {{h_{N_{{S_1}}^{'}k}}} \!\right|^2} \nonumber \\
&&\!\!\!\!\!\!\!\!\!\!\!\!\!\!\!\int_0^\infty\!\!\!\!\!\!  {\cdots \!\!\int_0^\infty\!\!\!\!\!  {f\!\left(\! {{{\left|\! {h_{1k}^t} \!\right|}^2}} \!\right)\!\cdots\!} } f\!\left(\! {{{\left|\! {h_{{N_{\rm{BS}}}k}^t} \!\right|}^2}} \!\right)d{\left|\! {h_{{N_{1k}}}^t} \!\right|^2}\!\cdots\! d{\left|\! {h_{{N_{\rm{BS}}}k}^t} \!\right|^2}\nonumber \\
&&\int_0^u {{\alpha _{{{\rm{N1}}_n}}}{e^{ - \left( {{\beta _{{{\rm{N1}}_n}}} - {\delta _{{{{\rm{N1}}_n}}}}} \right)x}}\sum\limits_{{n_{{{\rm{N1}}_n}}} = 0}^{{m_{{{\rm{N1}}_n}}} - 1}  {{{Z0}_{{{\rm{N1}}_n}}}dx} } \nonumber \\
&&\!\!\!\!\!\!\!\!\!\!\!\!\!\!=\!\! \int_0^\infty  \!\!\!\!\!{\!\cdots \! \int_0^\infty\!\!\!\!\!  {f\!\!\left(\!\! {{{\left| {{h_{1k}}} \right|}^2}} \!\!\right)\!\cdots\!} } f\!\!\left(\!\! {{{\left| {{h_{N_{{S_1}}^{'}k}}} \right|}^2}} \!\!\right)d{\left|\! {{h_{1k}}} \!\right|^2} \!\cdots\! d{\left|\! {{h_{N_{{S_1}}^{'}k}}} \!\right|^2} \nonumber \\
&&\!\!\!\!\!\!\!\!\!\!\!\!\!\!\int_0^\infty\!\!\!\!\!\!  {\cdots \!\!\int_0^\infty\!\!\!\!\!  {f\!\left(\! {{{\left|\! {h_{1k}^t} \!\right|}^2}} \!\right)\!\cdots\!} } f\!\left(\! {{{\left|\! {h_{{N_{\rm{BS}}}k}^t} \!\right|}^2}} \!\right)d{\left|\! {h_{{N_{1k}}}^t} \!\right|^2}\!\cdots\! d{\left|\! {h_{{N_{\rm{BS}}}k}^t} \!\right|^2} \nonumber \\
&&\sum\limits_{{n_{{\rm{N1}}_n}}=0}^{{m_{{\rm{N1}}_n}} \!-\! 1}\!\! {{Z_{{\rm{N1}}_n}}\!\!\left(\!\!
1 \!-\! {e^{ -\! \left(\! {{\beta _{{{\rm{N1}}_n}}} \!\!-\! {\delta _{{{{\rm{N1}}_n}}}}} \!\right)\!\!\sum\limits_{m = 1}^{N_{{\rm{S1}}}^{'}}\!\! {{\zeta _m}{{\left|\! {{h_{mk}}} \!\right|}^2}} }} \cdot \right.} \nonumber\\
&&{e^{ - \left( {{\beta _{{{\rm{N1}}_n}}} \!-\! {\delta _{{{{\rm{N1}}_n}}}}} \!\right)\!\!\sum\limits_{i \in {\phi _{\rm{BS}}}} \!\!{{I_{{\rm{BS}}_i}}{{\left| {h_{ik}^t} \right|}^2}} }}{e^{ -\! \left(\! {{\beta _{{{\rm{N1}}_n}}} \!-\! {\delta _{{{{\rm{N1}}_n}}}}} \!\right)S\sigma _{{\rm{N1}}}^2}}\nonumber \\
&&{\left.\sum\limits_{{k_o} \!=\! 0}^{{n_{{\rm{N1}}_n}}} \!\!\!{\frac{1}{{{k_o}!}}\frac{{{{\left(\!\! {\sum\limits_{m \!=\! 1}^{N_{{\rm{S1}}}^{'}}\!\! {{\zeta _m}{{\left|\! {{h_{mk}}} \!\right|}^2}}\!\!  +\!\!\!\!\!\! \sum\limits_{i \in {\phi _{\rm{BS}}}}\!\!\!\!\! {{I_{{\rm{BS}}_i}}{{\left|\! {h_{ik}^t} \!\right|}^2}}\!\!\!  +\!\! S\sigma _{{\rm{N1}}}^2} \!\!\right)}^{{k_o}}}}}{{{{\left( {{\beta _{{{\rm{N1}}_n}}} \!-\! {\delta _{{{{\rm{N1}}_n}}}}} \right)}^{ - {k_o}}}}}}\!\!\right)},
\end{eqnarray}
where ${{Z0}_{{{\rm{N1}}_n}}}=\frac{{( {{m_{{{\rm{N1}}_n}}} - 1} )!{{(\! {{\delta _{{{\rm{N1}}_n}}}x} \!)}^{{n_{{{{\rm{N1}}_n}}}}}}}}{{(\! {{m_{{{\rm{N1}}_n}}} \!-\! 1 \!-\! {n_{{{\rm{N1}}_n}}}} \!)!{n_{{{\rm{N1}}_n}}}!{{(\! 1 \!)}_{{n_{{{\rm{N1}}_n}}}}}}}$, and ${Z_{{{\rm{N1}}_n}}} = \frac{{{\alpha _{{{\rm{N1}}_n}}}( {{m_{{{{\rm{N1}}_n}}}} - 1} )!{{( {{\delta _{{{{\rm{N1}}_n}}}}} )}^{{n_{{{{\rm{N1}}_n}}}}}}}}{{( {{m_{{{{\rm{N1}}_n}}}} - 1 - {n_{{{{\rm{N1}}_n}}}}} )!{{( 1 )}_{{n_{{{{\rm{N1}}_n}}}}}}{{( {{\beta _{{{\rm{N1}}_n}}} - {\delta _{{{{\rm{N1}}_n}}}}} )}^{{n_{{{\rm{N1}}_n}}} + 1}}}}$. ${\sum\limits_{{k_o} = 0}^{{n_{{{\rm{N1}}_n}}}} \!\!\!{\frac{1}{{{k_o}!}}( {{\beta _{{{\rm{N1}}_n}}} \!\!-\!\! {\delta _{{{{\rm{N1}}_n}}}}} )} ^{{k_o}}}{( {\sum\limits_{m = 1}^{N_{{\rm{S1}}}^{'}}\!\!\! {{\zeta _m}{{| {{h_{mk}}} |}^2}}  \!\!+\!\!\!\!\! \sum\limits_{i \in {\phi _{BS}}}\!\!\!\! {{I_{{\rm{BS}}_i}}{{| {h_{ik}^t} |}^2}} \!\! + s\sigma _{{\rm{N1}}}^2} )^{{k_o}}}$ can be given by
\begin{eqnarray}
&&\!\!\!\!\!\!\!\!\!\!\!\!\!\!\!\!\!\!\!{\sum\limits_{{k_o} = 0}^{{n_{{{\rm{N1}}_n}}}}\!\! k_{\beta\delta}}\!{\left(\! {\sum\limits_{m = 1}^{N_{{\rm{S1}}}^{'}} \!{{\zeta _m}{{\left|\! {{h_{mk}}} \right|}^2}} \! +\! \sum\limits_{i \in {\phi _{BS}}} \!\!{{I_{{\rm{BS}}_i}}\!{{\left| {h_{ik}^t} \right|}^2}} \! +\! S\sigma _{{\rm{N1}}}^2} \!\right)^{{k_o}}}\nonumber\\
&&\!\!\!\!\!\!\!\!\!\!\!\!\!\!\!\!\!\!\!=\! \sum\limits_{{k_o} = 0}^{{n_{{\rm{N}}{1_n}}}}\! {\sum\limits_{{j_o} = 0}^{{k_o}}\! {\sum\limits_{{z_o=0}}^{{j_o}}\! k_{\beta\delta} } }
\left( {\begin{matrix}
   {{k_o}}  \\
   {{j_o}}  \\
 \end{matrix} } \right)\!\left( {\begin{matrix}
   {{j_o}}  \\
   {{z_o}}  \\
 \end{matrix} } \right){{\left( {S\sigma _{{\rm{N}}1}^2} \right)}^{{j_o} - {z_o}}}\nonumber\\
&&\!\!\!\!\!\!\!\!\!\!\!\!\! \sum\limits_{x_m\ge0,{x_1},{x_2},\cdots,{x_{N_{{\rm{S1}}}^{'}}}}^{{k_o-j_o}} \!\!\!\!\!\!\!\!\!{k_{jx}} {\left(\! {{\zeta _1}{{\left| {{h_{1k}}} \right|}^2}} \!\right)^{{x_1}}}\!\!\cdots{\left(\! {{\zeta _{N_{{\rm{S1}}}^{'}}}{{\left| {{h_{N_{{\rm{S1}}}^{'}k}}} \right|}^2}} \!\right)^{{x_{N_{{\rm{S1}}}^{'}}}}} \nonumber \\
&&\!\!\!\!\!\!\!\!\!\!\!\!\!\!\!\!\!\!\!\sum\limits_{y_i\ge0,{y_1},{y_2},\cdots,{y_{{N_{\rm{BS}}}}}}^{{z_o}} \!\!\!\!\!\!\!\!\!\!\!\!\!\!{z_{y}} {\left(\! {{I_{{{\rm{BS}}_1}}}{{\left|\! {h_{1k}^t} \!\right|}^2}} \!\right)^{{y_1}}}\!\!\cdots{\left(\! {{I_{{\rm{BS}}_{{N_{\rm{BS}}}}}}{{\left|\! {h_{{N_{\rm{BS}}}k}^t} \!\right|}^2}} \!\right)^{{y_{{N_{\rm{BS}}}}}}},
\end{eqnarray}
where $k_{\beta\delta}={\frac{1}{{{k_o}!}}( {{\beta _{{{\rm{N1}}_n}}} - {\delta _{{{{\rm{N1}}_n}}}}} )^{{k_o}}}$, $k_{jx} = ( {\begin{matrix}
   {{k_o-j_o}}  \\
   {{x_1},{x_2},\cdots,{x_{N_{{\rm{S1}}}^{'}}}}  \\
 \end{matrix}} )$, and $z_{y}=( {\begin{matrix}
   {{z_o}}  \\
   {{y_1},{y_2},\cdots,{y_{{N_{\rm{BS}}}}}}  \\
 \end{matrix} } )$.

Upon relying on \cite{BIB30}, and substituting (A.6) into (A.5), we have
\begin{eqnarray}
&&\!\!\!\!\!\!\!\!\!\!\!\!\!\!\!\!\!\!\!\!\!\!\!\!P_{{\rm{OP}}}^{{\rm{NGSO1}}}= \sum\limits_{{n_{{\rm{N}}{1_n} = 0}}}^{{m_{{\rm{N}}{1_n} \!-\! 1}}} \!\!{\left(\!\! {{Z_{{\rm{N}}{1_n}}}\!\!\! \prod\limits_{m = 1}^{{{N}}_{{\rm{S1}}}^{'}} \sum\limits_{{n_{{\rm{N}}{2_m}}} = 0}^{{m_{{\rm{N}}{2_m}}}\! - \!1}\!\! {Z_{{{\rm{N2}}_m}}}} \right.}-  \nonumber \\
&& {Z_{{\rm{N}}_{{1_n}}}}\!\!{e^{ -\! \left(\! {{\beta _{{{\rm{N1}}_n}}} \!-\! {\delta _{{{{\rm{N1}}_n}}}}} \!\right)S\sigma _{{\rm{N1}}}^2}}\!\sum\limits_{{k_o} = 0}^{{n_{{\rm{N}}{1_n}}}}\! {\sum\limits_{{j_o} = 0}^{{k_o}}\! {\sum\limits_{{z_o=0}}^{{j_o}} {k_{\beta\delta}} \left( {\begin{matrix}
   {{k_o}}  \\
   {{j_o}}  \\
 \end{matrix} } \right)} } \nonumber\\
&&\left( {\begin{matrix}
   {{j_o}}  \\
   {{z_o}}  \\
 \end{matrix} } \right){{\left( {S\sigma _{{\rm{N}}1}^2} \right)}^{{j_o} \!-\! {z_o}}}\!\sum\limits_{x_m\ge0,{x_1},{x_2},...,{x_{{{N}}_{{\rm{S1}}}^{'}}}}^{{k_o\!-\!j_o}}\!\!\!\!\!\!\!\!{{k_{jx}}\!\!\prod\limits_{m = 1}^{{{N}}_{{\rm{S1}}}^{'}}\! {{{\left( {{\zeta _m}} \right)}^{{x_m}}}} }\nonumber\\
&& \prod\limits_{m = 1}^{{{N}}_{{\rm{S1}}}^{'}} \!{\!\left(\! {\sum\limits_{{n_{{\rm{N}}{2_m}}}=0}^{{m_{{\rm{N}}{2_m}}} \!-\! 1}\! {{{Z}_{{{\rm{N2}}_m}}}}\left({{\beta _{{{\rm{N}}_{{2_m}}}}}\! -\! {\delta _{{{{\rm{N}}_{{2_m}}}}}}}\right) \left( {{x_m} \!+\! {n_{{\rm{N}}{2_m}}}} \right)!} \right.} \nonumber\\
&&\;\;\;\;\;\;{\!\left(\! {\left(\! {{\beta _{{{\rm{N}}_{{1_n}}}}}\! -\! {\delta _{{{{\rm{N}}_{{1_n}}}}}}} \!\right)\!{\zeta _m}{{ \!+\! }}\left(\! {{\beta _{{\rm{N}}{2_m}}} \!-\! {\delta _{{{\rm{N}}{2_m}}}}} \!\right)} \!\right)^{ - {x_m} \!-\! {n_{{\rm{N}}{2_m}}}\!-\!1}} \nonumber\\
&&\;\;\;\;\;\;\sum\limits_{y_i\ge 0,{y_1},{y_2},\cdots,{y_{{{\rm{N}}_{{\rm{BS}}}}}}}^{{z_o}} \!{{z_y}\!\prod\limits_{i = 1}^{{{\rm{N}}_{{\rm{BS}}}}}\! {{{\left( {{I_{{\rm{BS}}{_i}}}} \right)}^{{y_i}}}} } \!\prod\limits_{i = 1}^{{{\rm{N}}_{{\rm{BS}}}}} \! {\frac{1}{{\sigma _{{\rm{BS}}}^2}}\!\left( {{y_i}} \right)!}  \nonumber\\
&&\;\;\;\;\;\;\left. {\left. {{{\left( {\left( {{\beta _{{{\rm{N}}_{{1_n}}}}} - {\delta _{{{{\rm{N}}_{{1_n}}}}}}} \right){I_{{\rm{BS}}{_i}}} + \frac{1}{{\sigma _{{\rm{BS}}}^2}}} \right)}^{ - {y_i} - 1}}} \right)} \right).
\end{eqnarray}

\section{Derivation of ${P_{\rm{AOP}}^{\rm{BS}}}$}
Using (A.3), ${P_{\rm{AOP}}^{\rm{BS}}}$ can be written as
\begin{eqnarray}
&&\!\!\!\!\!\!\!\!\!\!\!{P_{\rm{AOP}}^{\rm{BS}}} = \Pr \left( {|h_{ik}^t{|^2} < S_{abcd}}\right)\nonumber\\
&&\!\!\!\!\!\!\!\!\!\!\! =\int_0^\infty\!\cdots  {\int_0^\infty\!\!\!\!  {f\!\left(\! {{{\left| {{h_{1k}}} \right|}^2}} \!\right)\!\cdots\!} } f\left(\! {{{\left| {{h_{{\rm{N}}_{{S_3}}^{'}k}}} \right|}^2}} \!\right)\!d{\left| {{h_{1k}}} \right|^2}\!\cdots \!d{\left| {{h_{{\rm{N}}_{{S_3}}^{'}k}}} \right|^2}\nonumber\\
&&\!\!\!\!\!\int_0^\infty\!\cdots  {\int_0^\infty\!\!\!\!\!\!  {f\!\left(\! {{{\left|\! {{h_{1k}}} \!\right|}^2}} \!\right)\!\cdots\!} } f\!\left(\! {{{\left|\! {{h_{{\rm{N}}_{{S_4}}^{'}k}}} \!\right|}^2}} \!\right)d{\left|\! {{h_{1k}}} \!\right|^2} \!\cdots\! d{\left|\! {{h_{{\rm{N}}_{{S_4}}^{'}k}}} \!\right|^2}\nonumber\\
&&\!\!\!\!\!\int_0^\infty\!\cdots  {\int_0^\infty\!\!\!\!\!\!  {f\!\left(\! {{{\left|\! {h_{1k}^t} \!\right|}^2}} \!\right)\!\cdots\!} } f\!\left(\! {{{\left|\! {h_{{{\rm{N}}_{{\rm{BS}}}}k}^t} \!\right|}^2}} \!\right)\left(\!\! {1 \!-\! {e^{ - \frac{1}{{\sigma _{{\rm{BS}}}^2}}S_{abcd}}}} \!\!\right)\nonumber\\
&&\;\;\;\;\;\;\;\;\;\;\; d{\left|\! {h_{{{\rm{N}}_{1k}}}^t} \!\right|^2}\!\cdots\! d{\left|\! {h_{{{\rm{N}}_{{\rm{BS}}}}k}^t} \!\right|^2},
\end{eqnarray}
where $A_n^{'} = {A_n}{P_i}$, $B_m^{'} = {B_m}{P_i}$, $C_j^{'} = {C_j}{P_i}$, $D_i^{'} = {D_i}{P_i}$, and $S_{abcd} = \sum\limits_{n = 1}^{{{N}}_{{\rm{S3}}}^{'}} {{A_n}\!{{| {{h_{nt}}} |}^2}} \! +\! \sum\limits_{m = 1}^{{{N}}_{{\rm{S4}}}^{'}} {{B_m}{{| {{h_{mt}}} |}^2}}  \!+\!\! \sum\limits_{j = 1,j \ne i}^{{{\rm{N}}_{{\rm{BS}}}}} {{C_j}{{| {h_{jt}^t} |}^2}}  + {D_i}$.

When ${P_i} \to \infty $, then $1 - {e^{ - \frac{1}{{\sigma _{{\rm{BS}}}^2{P_i}}}S_{abcd}}}$ can be given by
\begin{equation}
1 - {e^{ - \frac{1}{{\sigma _{{\rm{BS}}}^2{P_i}}}S_{abcd}}}= \frac{1}{{\sigma _{{\rm{BS}}}^2{P_i}}}S_{abcd},
\end{equation}

Furthermore, ${P_{\rm{AOP}}^{\rm{BS}}}$ can be finally formulated as
\begin{eqnarray}
&&\!\!\!\!\!\!\!\!\!\!\!P_{\rm{AOP}}^{\rm{BS}} = \frac{1}{{\sigma _{{\rm{BS}}}^2\!{P_i}}}\!D_i^{'}\!\prod\limits_{n = 1}^{{\rm{N}}_{{S_3}}^{'}} \sum\limits_{{n_{{\rm{N}}{1_n}}} = 0}^{{m_{{\rm{N}}{1_n}}} \!-\! 1}{ {Z_{{{\rm{N1}}_n}}} } \prod\limits_{m = 1}^{{{N}}_{{\rm{S4}}}^{'}}\! {\sum\limits_{{n_{{\rm{N}}{2_m}}} \!=\! 0}^{{m_{{\rm{N}}{2_m}}} \!-\! 1}\! {Z_{{{\rm{N2}}_m}}} } +\nonumber\\
&&\!\!\!\!\!\!\!\!\!\!\!\frac{1}{{\sigma _{{\rm{BS}}}^2\!{P_i}}}\!\prod\limits_{n \!=\! 1}^{{\rm{N}}_{{S_3}}^{'}} \!{\sum\limits_{{n_{{\rm{N}}{1_n}}} \!=\! 0}^{{m_{{\rm{N}}{1_n}}} \!-\! 1} \!\!\!\!{Z_{{{\rm{N1}}_n}}} } \prod\limits_{m \!= \!1}^{{{N}}_{{\rm{S4}}}^{'}}\! {\sum\limits_{{n_{{\rm{N}}{2_m}}} \!=\! 0}^{{m_{{\rm{N}}{2_m}}} \!-\! 1} \!\!\!\!{Z_{{{\rm{N2}}_m}}} }\!\!\!\! {\prod\limits_{j = 1,j \ne i}^{{{\rm{N}}_{{\rm{BS}}}}}\!\!\!\! {C_j^{'}\left( {\frac{1}{{\sigma _{{\rm{BS}}}^2}}} \right)} ^{ - 2}}+\nonumber\\
&&\!\!\!\!\!\!\!\!\!\!\!\frac{1}{{\sigma _{{\rm{BS}}}^2\!{P_i}}}\!\prod\limits_{n = 1}^{{\rm{N}}_{{S_3}}^{'}} \!{\sum\limits_{{n_{{\rm{N}}{1_n}}} \!=\! 0}^{{m_{{\rm{N}}{1_n}}}\! -\! 1} \!{Z_{{{\rm{N1}}_n}}} }\prod\limits_{m = 1}^{{{N}}_{{\rm{S4}}}^{'}}\! {B_m^{'}\!\sum\limits_{{n_{{\rm{N}}{2_m}}} \!=\! 0}^{{m_{{\rm{N}}{2_m}}} \!-\! 1} }\! \frac{Z_{{{\rm{N2}}_m}}}{\left(\! {{\beta _{{\rm{N}}{2_m}}} \!-\! {\delta _{{\rm{N}}{2_m}}}} \!\right)}+\nonumber \\
&&\!\!\!\!\!\!\!\!\!\!\!  \frac{1}{{\sigma _{{\rm{BS}}}^2{P_i}}}\prod\limits_{m = 1}^{{{N}}_{{\rm{S4}}}^{'}} {\sum\limits_{{n_{{\rm{N}}{2_m}}} = 0}^{{m_{{\rm{N}}{2_m}}} - 1} \!\!\!\!{Z_{{{\rm{N2}}_m}}} }\prod\limits_{n = 1}^{{\rm{N}}_{{S_3}}^{'}}A_n^{'} {\sum\limits_{{n_{{\rm{N}}{1_n}}} \!=\! 0}^{{m_{{\rm{N}}{1_n}}} \!-\! 1} }\!\!\!\!{Z_{{{\rm{N1}}_n}}}{{\left( {{n_{{\rm{N}}{1_n}}} + 1} \right)!} }\nonumber\\
&&\;\;\;\;\;\;\;\;\;\;\;\;\;\;\;\;\;\;\;\;\;\;\;\;\;\;\;\;\;\;\;\;\;\;\;\;\;\;\;\;\;\;\;\; {\left( {{\beta _{{\rm{N}}{1_n}}} - {\delta _{{\rm{N}}{1_n}}}} \right)^{ - 1}}.
\end{eqnarray}
\ifCLASSOPTIONcaptionsoff
  \newpage
\fi

%\end{spacing}
\end{document}